\documentclass[preprint,review,10pt,number,sort&compress]{elsarticle}
\usepackage{setspace}
\usepackage[margin=3cm]{geometry}

\usepackage{graphicx}
\usepackage{amssymb}
\usepackage{amsmath}
\usepackage{lineno}
\usepackage{subfigure}
\usepackage{xcolor}
\usepackage{soul}
\usepackage{graphicx}
\usepackage{placeins}
\usepackage{import}

\newcommand{\bx}{\mathbf{x}}
\newcommand{\by}{\mathbf{y}}
\newcommand{\tbx}{\tilde{\bx}}
\newcommand{\tby}{\tilde{\by}}
\newcommand{\epsexp}{\varepsilon_{\text{exp}}}
\newcommand{\epspred}{\varepsilon_{\text{pred}}}
\newcommand{\sigmaexp}{\sigma_{\text{exp}}}
\newcommand{\tauexp}{\tau_{\text{exp}}}
\newcommand{\GP}{\operatorname{GP}}

\newcommand{\qref}[1]{Eq.~(\ref{#1})}
\newcommand{\fref}[1]{Fig.~\ref{#1}}
\newcommand{\sref}[1]{Sec.~\ref{#1}}

\usepackage[obeyspaces]{url}

\usepackage{tikz}
\usetikzlibrary{bayesnet}
\usepackage{xcolor}

\begin{document}
\begin{frontmatter}

%% Title
\title{Bayesian Inference of Fiber Orientation and Polymer Properties in Short Fiber-Reinforced Polymer Composites}

%% Date

% Author
\author[AERO,CMSC]{Akshay J.Thomas \corref{cor1}}
\ead{ajacobth@purdue.edu}
\cortext[cor1]{Corresponding Author}
\address[AERO]{School of Aeronautics and Astronautics, Purdue University, West Lafayette, IN, 47907}
\address[CMSC]{Composites Manufacturing and Simulation Center, Purdue University, West Lafayette, IN 47906}
\address[MECH]{School of Mechanical Engineering, Purdue University, West Lafayette, IN, 47907}

\author[CMSC]{Eduardo Barocio}
\author[MECH]{Ilias Bilionis}
\author[AERO,CMSC] {R. Byron Pipes}

\begin{abstract}
We present a Bayesian methodology to infer the elastic modulus of the constituent polymer and the fiber orientation state in a short-fiber reinforced polymer composite (SFRP). 
The properties are inversely determined using only a few experimental tests.
Developing composite manufacturing digital twins for SFRP composite processes, including injection molding and extrusion deposition additive manufacturing (EDAM) requires extensive experimental material characterization.
In particular, characterizing the composite mechanical properties is time consuming and therefore, micromechanics models are used to fully identify the elasticity tensor. 
Hence, the objective of this paper is to infer the fiber orientation and the effective polymer modulus and therefore, identify the elasticity tensor of the composite with minimal experimental tests.
To that end, we develop a hierarchical Bayesian model coupled with a micromechanics model to infer the fiber orientation and the polymer elastic modulus simultaneously which we then use to estimate the composite elasticity tensor. 
We motivate and demonstrate the methodology for the EDAM process but the development is such that it is applicable to other SFRP composites processed via other methods. 
Our results demonstrate that the approach provides a reliable framework for the inference, with as few as three tensile tests, while accounting for epistemic and aleatory uncertainty.
Posterior predictive checks show that the model is able to recreate the experimental data well.
The ability of the Bayesian approach to calibrate the material properties and its associated uncertainties, make it a promising tool for enabling a probabilistic predictive framework for composites manufacturing digital twins.

\end{abstract}

\journal{arxiv}
\end{frontmatter}

%

%% main text
\section{Introduction}
\label{se:intro}

Short-fiber reinforced polymer composites (SFRPs) are manufactured by injection molding or extrusion deposition of discontinuous fiber-filled thermoplastic pelletized polymers~\cite{tseng2017numerical}. 
SFRPs are used in lightweight applications owing to the increased stiffness and strength provided by the fibers~\cite{yashiro2012particle, yashiro2011numerical}. 
Extrusion Deposition Additive Manufacturing (EDAM) is a process by which fiber-filled polymer systems are extruded in a screw and deposited onto a build platform in a path dictated by a set of machine instructions generally in the form of a machine code. 
These fiber reinforced polymers have allowed the printing of geometries in the scale of meters by increasing the stiffness and lowering the thermal expansion of the printed material, dominantly in the printing direction~\cite{love2014importance}. 
The application of this method include printing geometries for autoclave tooling and compression molding~\cite{barocio2017extrusion, hassen2016durability}.
The printing process is a multi-physics problem involving anisotropic heat transfer, shrinkage, creep and stress relaxation, polymer crystallization and melting,and fusion  bonding. 
The effects of these phenomena manifest as deformation and residual stresses that can cause issues such as delamination. 
This has motivated the development of physics-based~\cite{brenken2019development, Barocio2020, hebert2016holistic, talagani2015numerical} methods to predict the outcome of the printing process.
To reduce the time and costly empirical calibration of processing parameters, there has been much interest in developing simulation capabilities to predict the outcomes of a given print geometry or strategy~\cite{brenken2019development, hebert2016holistic}.
The goal of these studies has been to develop manufacturing digital twins that can predict the process induced deformation and residual stresses.
The mechanical properties of the composite need to be provided as inputs to the manufacturing simulations to predict the residual deformation and stress state.

Experimentally characterizing the anisotropic mechanical properties is extremely time consuming and therefore, micromechanics models are generally used to augment experimental measurements. 
To make use of suitable micromechanics models, we require information on microstructural descriptors such as the fiber orientation state, fiber length distribution, fiber volume fraction, and the mechanical properties of the constituent materials. 
The fiber orientation state is one of the key microstructural descriptors that governs the mechanical, transport, and thermoviscoelastic properties of the composite~\cite{mortazavian2015effects, ogierman2016study}.

The fiber orientation state can be experimentally measured using computed tomography (CT) scans and optical microscopy.
However, CT scanning is expensive whereas optical microscopy is a destructive technique that requires a sample sectioned from a printed geometry.
Further, resolving fiber orientation from 2D images is restricted to cylindrical fibers~\cite{bay1992stereological} or clusters of non-cylindrical fibers~\cite{sharp2019measuring}.
Similarly, CT scans are restricted to volumes that are smaller than the volume of a printed bead.
Also, these methods are prone to human error and require multiple iterations of the same experiment to obtain reliable data. 
While simulation methods have been developed to predict the fiber orientation state in fused filament fabrication (FFF)~\cite{papon2020review} and injection molding~\cite{tseng2017numerical, advani1987prediction, gupta1993fiber}, such methods are underdeveloped in EDAM. 
The complexity added due to the extruder and the gear pump makes the problem even more challenging. 
In addition to the difficulty of measuring fiber orientation, the processability of the polymer is modified via additives which in turn can alter its elastic properties. 
In light of the difficulties posed by experimental measurements, past work has used inverse determination approaches to calibrate numerical models.

Digimat~\cite{adam2014integrated} provides a suite of deterministic ``reverse engineering'' (inverse determination) tools to calibrate the properties of the material constituents using experimental data. 
Digimat currently provides out-of-the-box capabilities to infer the parameters of pre-defined material models, e.g., elastic properties of a constituent phase in a multiphase system, $J_2$ plasticity  and viscoplasticity model parameters, viscoelastic model parameters, etc.
Digimat can also be used to infer the average aspect ratio of an inclusion from experimental data. 
Calibration of numerical models is typically achieved my minimizing the error between a vector of experimentally measured quantities and a vector of model predictions. 
This is commonly achieved via the least squares method. 
This class of problems is ubiquitous in the medical imaging~\cite{arridge1997optical, arridge1999optical},  and geophysics domains~\cite{menke2018geophysical, russell1988introduction}. 
Kurkin \textit{et al.} calibrated a non-linear material model for SFRPs using Digimat's inverse determination tool~\cite{kurkin2020influence, kurkin2020mechanical}. 
Landervik \textit{et al.} used this methodology to calibrate an elastoplastic material model~\cite{landervik2015digimat} . 
\citet{lindhult2015fatigue} used Digimat to calibrate a pseudo-grain-based fatigue model.
The least-squares approach, however, results in a deterministic estimate of the model parameters. 
To overcome this limitation, and account for the uncertainty in the calibrated model parameters, researchers have used a Bayesian inference (BI) framework.
 
 BI has been used to statistically calibrate elasticity constants of glass-fiber composites~\cite{lai1996parameter} and carbon fiber composites~\cite{daghia2007estimation} through vibration tests. 
Similarly, ~\citet{gogu2013bayesian} used BI to infer the elastic constants of a graphite-epoxy composites through open-hole testing.
\citet{albuquerque2018bayesian} calibrated cohesive zone models and~\citet{sankararaman2011uncertainty} calibrated fatigue crack growth models using BI.
Further reported applications of BI include calibration of elasto-plastic material model parameters~\cite{rappel2019identifying, rappel2020tutorial}, visco-elastic model parameters~\cite{rappel2018bayesian}, and hyperelasticity models~\cite{madireddy2015bayesian}.
The work by Mohamedou~\cite{mohamedou2019bayesian} is the most relevant to the work presented in this paper. 
They used BI to calibrate the matrix modulus and the fiber aspect ratio. 
They present two inference methods - ``error-based inference'' and ``distribution-based inference.'' 
The ``error-based inference'' was based on experimentally measured composite moduli values. 
They model the moduli measurement process using a Gaussian distribution with a fixed standard deviation obtained from the experimental measurements.
The parameters inferred using this method were not able to recreate the experimental stress-strain data within a 95\% credible interval.
This could be due to the fact that the Gaussian measurement model was not suitable in this case.  
In the ``distribution-based inference'' method only the polymer modulus was inferred. 
They kept the aspect ratio a constant and chose its value from the previously discussed ``error-based inference.''
For the ``distribution-based inference'' method, they modeled the measurement process using a Beta distribution. 
For each composite moduli measured, the corresponding polymer modulus was calculated by inverting the micromechanics model to form a vector of the so-called experimentally measured ``polymer modulus.'' 
The inferred parameters using this approach predicted a 95\% credible interval which covers all the measured experimental data.
\citet{wu2020bayesian} extended this framework to calibrate the uncertainties in non-linear micromechanics model parameters.
Though these methods provide a reliable and comprehensive method to infer the aspect ratio and effective polymer properties, there still remains the question of whether we can infer the fiber orientation from similar measurements.
 
Despite the advancement in simulation capabilities to predict manufacturing process induced fiber orientation in injection and compression molding, similar models are underdeveloped in EDAM.
Further, to the best of the authors' knowledge, existing literature has been unable to infer the orientation state of the fibers and the modified properties of the polymer using inverse determination methods.
Hence, the objective of this paper is to address the latter limitation existing in the state-of-the-art. 
We introduce a methodology that enables the inference of the orientation of the fibers and the modified effective polymer properties. 
We demonstrate the method for the EDAM process in particular to elaborate the approach but the development is such that it is applicable to other SFRP systems. 
We present a hierarchical Bayesian methodology that uses limited experimental data and a suitable micromechanics model to characterize the anisotropic mechanical properties of the composite.
We propose a three component Bayesian inference process.
First, we utilize a micromechanics model to predict the composite elastic properties from the constituent properties. 
Second, we conduct three sets of experimental tests to infer the fiber orientation state and the effective polymer modulus. And, third, we implement a hierarchical Bayesian framework for the inference. We discuss each of these components in the following sections.

This paper is organized as follows. We start by discussing the micromechanics model used and experimental measurements in \sref{se:micro_model}. 
Next, we present the Bayesian inference model in \sref{se:BI} and the Gaussian Process surrogate to replace the micromechanics model for computational feasibility in \sref{se:GPR}. The results and discussions are presented next in \sref{se:res_and_dis}. Finally, we present our conclusions and discuss avenues for future work in \sref{se:conc}.

%-----------------------------------Micromechanics and experiments ----------------------------%

\section{Micromechanics Model and Experimental Methods}
\label{se:micro_model}

\subsection{Micromechanics Model}

First, we discuss the micromechanics model. Multiphase systems are typically homogenized using a Mean Field Homogenization (MFH) method.
In this work, we use a two-step homogenization method~\cite{doghri2005micromechanical} (implemented in Digimat), which uses the Mori-Tanaka~\cite{mori1973average} model.
The model invokes the dilute suspension assumption making it suitable for composites with low fiber volume fraction, upto approximately 25\% by weight for carbon fibers. 
Therefore, this work focuses on low volume fraction composite materials. 
Nevertheless, it is important to highlight that the methodology discussed is applicable to higher fiber content using a suitable micromechanics model that represents the high fiber content system well.

\begin{figure}[h]
  \centering
  \includegraphics[width=0.7\linewidth]{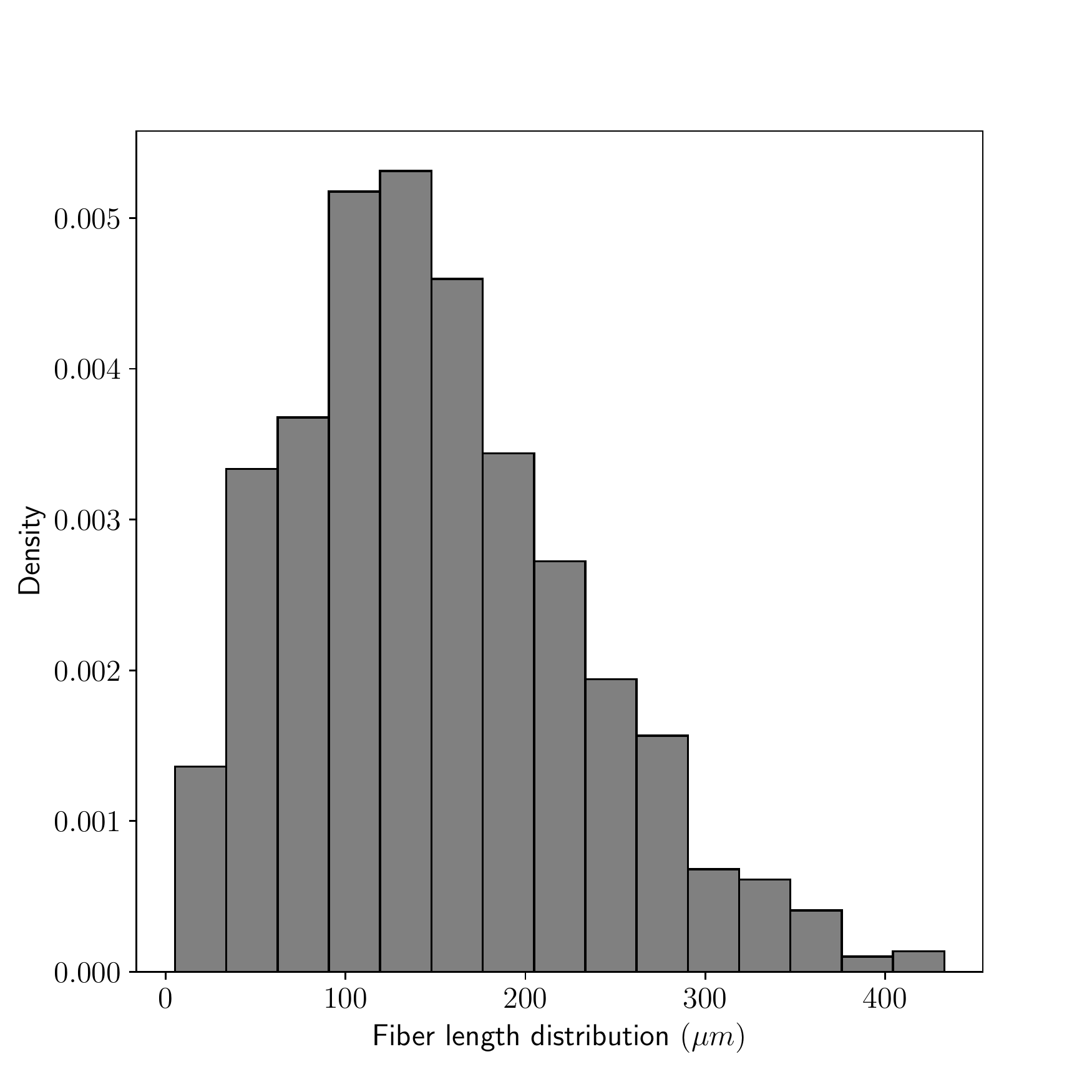} % scale=0.6
  \caption{Fiber length distribution of the processed 20 percent by weight CF-PEI.
  }
  \label{fi:fiber_length}
\end{figure}

The micromechanics model requires a list of inputs in addition to the fiber orientation and the polymer elastic properties, which we aim to infer in this work. 
These include the mechanical properties of the transversely isotropic or isotropic (glass) fibers, the fiber aspect ratio, and the mass fraction of the fibers.
The material system used in this study was short fiber 20\% by weight carbon fiber (CF)- Polyethyleneimine (PEI) supplied by SABIC\textsuperscript{\textcopyright} (LNP\texttrademark THERMOCOMP\texttrademark AM COMPOUND EC004EXAR1).
We found the transversely isotropic fiber properties in literature~\cite{king1992micromechanics} and measured the fiber length distribution using optical microscopy after burning off the polymer and dispersing the fibers on a glass slide with silicone oil as outlined in~\cite{ramirez2018silico}.  
\fref{fi:fiber_length} shows the fiber length distribution measured experimentally.
We divided the fiber length by an average fiber diameter of 6 $\mu m$ to obtain a fiber aspect ratio distribution and use the fiber aspect ratio distribution in the two step MFH.
To describe the orientation state of the fibers, we use the second order orientation tensor, $\mathbf{A}$, following Advani and Tucker~\cite{advani1987use}.
The orientation tensor was assumed to possess only the diagonal terms, i.e., $\mathbf{A} = \mathrm{diag}(a_{11}, a_{22}, a_{33})$ as done in~\cite{brenken2019development}. Since the diagonal terms in the orientation tensor add to one, the properties to be inferred are matrix modulus $E$, matrix Poisson’s ratio ($\nu$), and the two terms of the orientation tensor ($a_{11}, a_{22}$). 
The variable $a_{33}$  is calculated by $a_{33}=1-a_{11}-a_{22}$.

\subsection{Experimental Measurements Of Composite Elastic Properties}
\begin{figure}[h!]

\subfigure[]{%
  \includegraphics[clip, width=0.95\columnwidth]{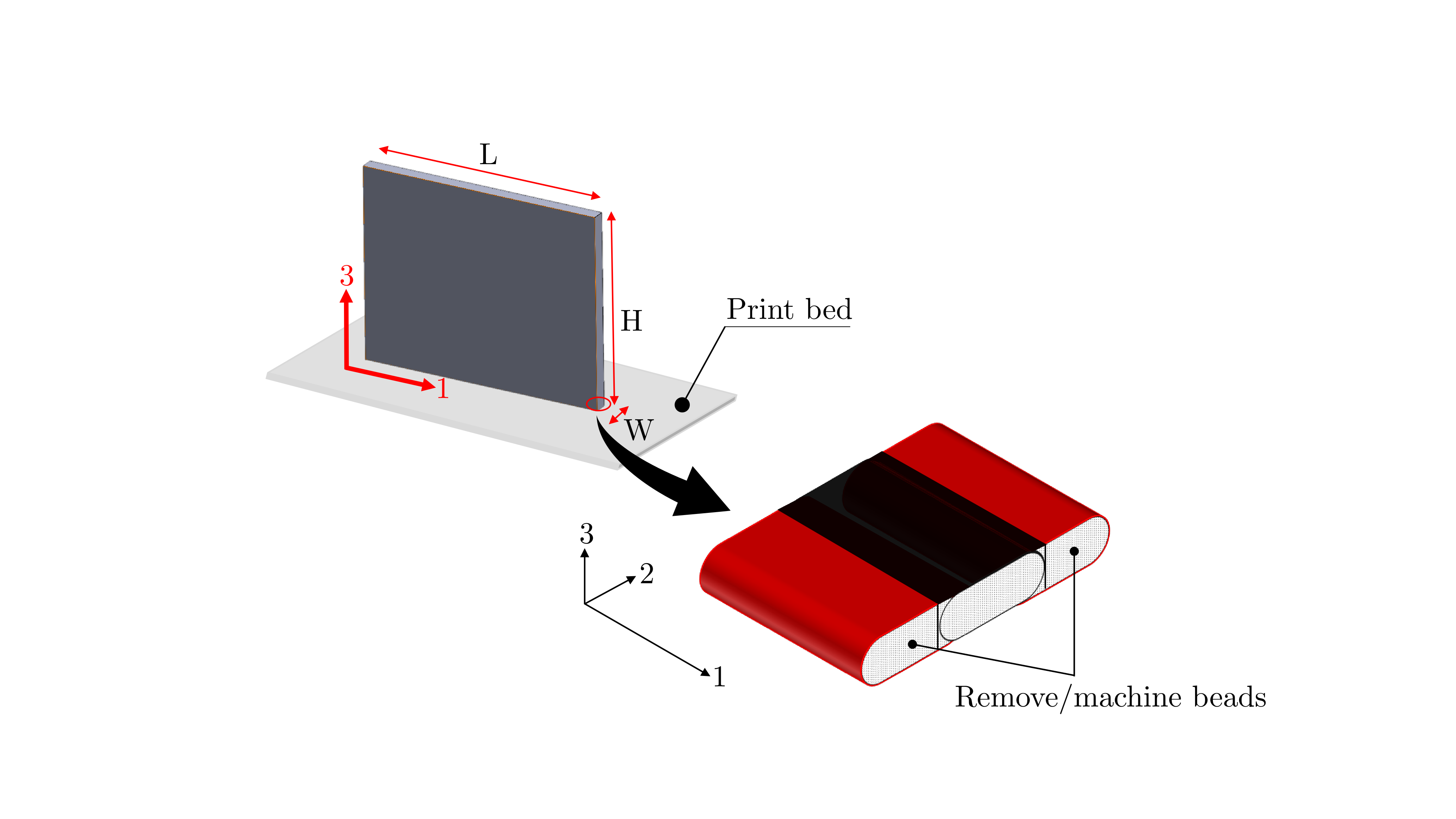}%clip, width=\columnwidth, scale = 0.4
}
\subfigure[]{%
  \includegraphics[clip, width=0.95\columnwidth]{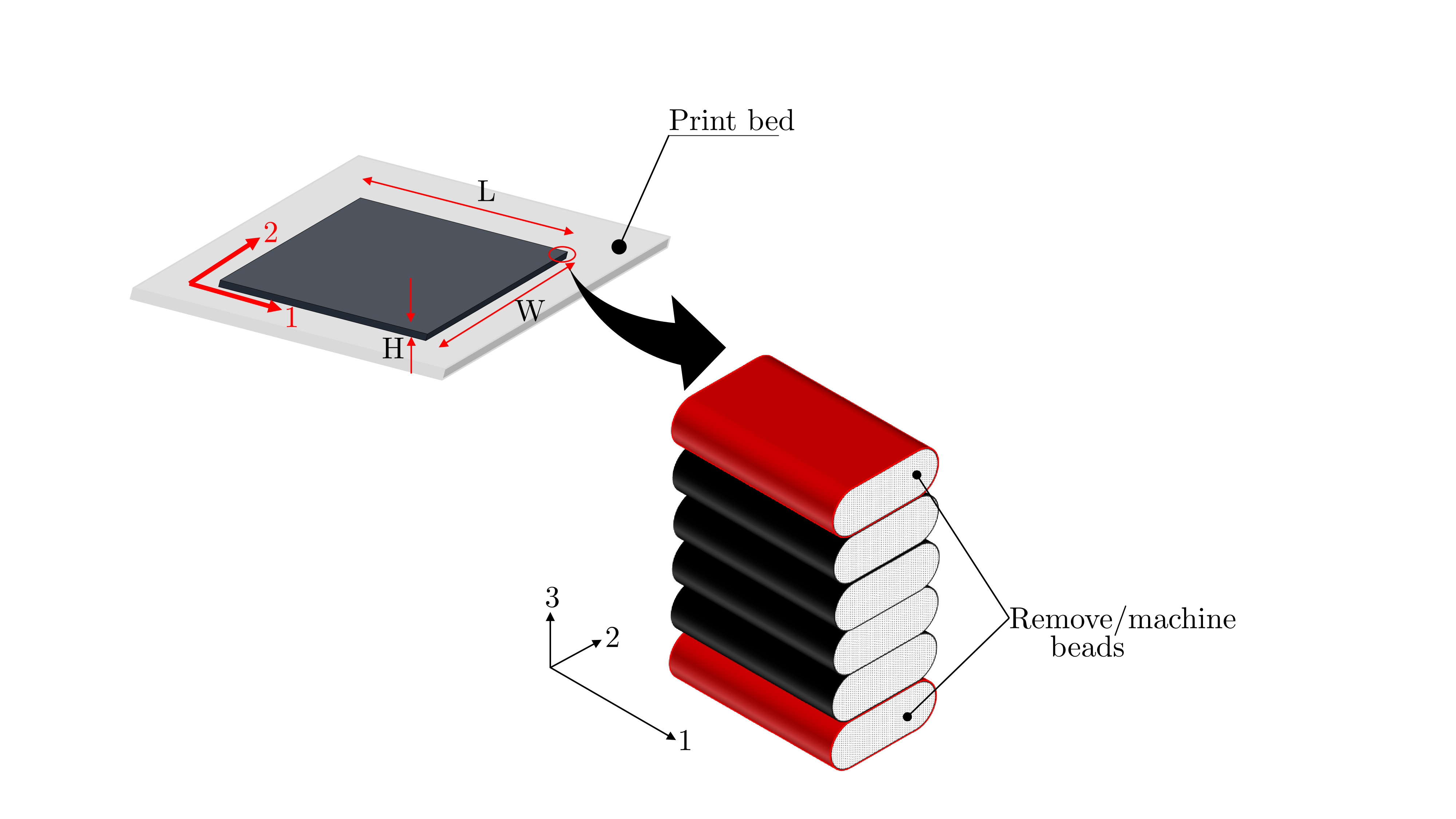}%clip, width=\columnwidth, scale = 0.4
}
\caption{Illustration of (a) Panel A and (b) Panel B used to prepare tensile specimens. Note that the beads highlighted in red are machined away.}
\label{fi:PanelAB}
\end{figure}

Having discussed the micromechanics model, we discuss the methods to prepare specimens and the experimental tests conducted. 
We require three sets of specimens to perform the necessary tests to obtain the tensile stress-strain in the three material directions. 
The 1-direction corresponds to the printing direction, the 2-direction corresponds to the direction transverse to this direction, and the 3-direction corresponds to the stacking direction of the printed beads. 
First, we describe the steps taken to prepare the specimen. 
We printed panels to extract coupons for the tensile tests using an extrusion deposition system developed at Purdue University, called the Composites Additive Manufacturing Research Instrument (CAMRI)~\cite{barocio2020fusion}.
The nozzle diameter of the extrusion system was 4 mm and the resulting dimensions of the printed bead were 6.15 mm in width and 1.5 mm in height. 
We printed two types of panels, printed in two different orientations using the same processing conditions, to extract specimens for tensile tests. 
Figure \ref{fi:PanelAB} (a) shows the panel used to prepare specimens for characterizing the elastic properties in the 1 and 3 directions. Similarly, \fref{fi:PanelAB} (b) shows the panel used to prepare specimens for characterizing the elastic properties in the 2 direction. 
From \fref{fi:PanelAB}(a) it is seen that the first panel, panel A, was printed three  beads  across  the  width  which  produced  panels  with final  dimensions  of  L=320  mm,  W  =  18.45  mm,  H  =  321  mm . The outermost beads were removed using a machining operation (shown in red in \fref{fi:PanelAB}).
The other panel, Panel B shown in \fref{fi:PanelAB}(b), was printed with six beads through the 3-direction with final dimensions of L= 175 mm, W = 159.9 mm, and H = 9 mm. The top and bottom beads in panel B were removed through a machining operation.
After the machining process, both panels were heat treated at the glass transition temperature of the polymer ($ 210$\textdegree C) to promote relaxation of thermal residual stresses. 
Next, we sectioned specimens for the tensile tests from the panels using a water jet and dried them in an oven 100 C for an hour to remove any moisture from the water-jetting process.
We spray painted the specimens with white matte paint and speckled them in a black stochastic pattern to prepare them for digital image correlation. 
Only a thin layer of white paint was applied to make sure the compliance of the paint was not measured using the DIC. 

\begin{figure}[h]
  \centering
  \includegraphics[width=0.9\linewidth]{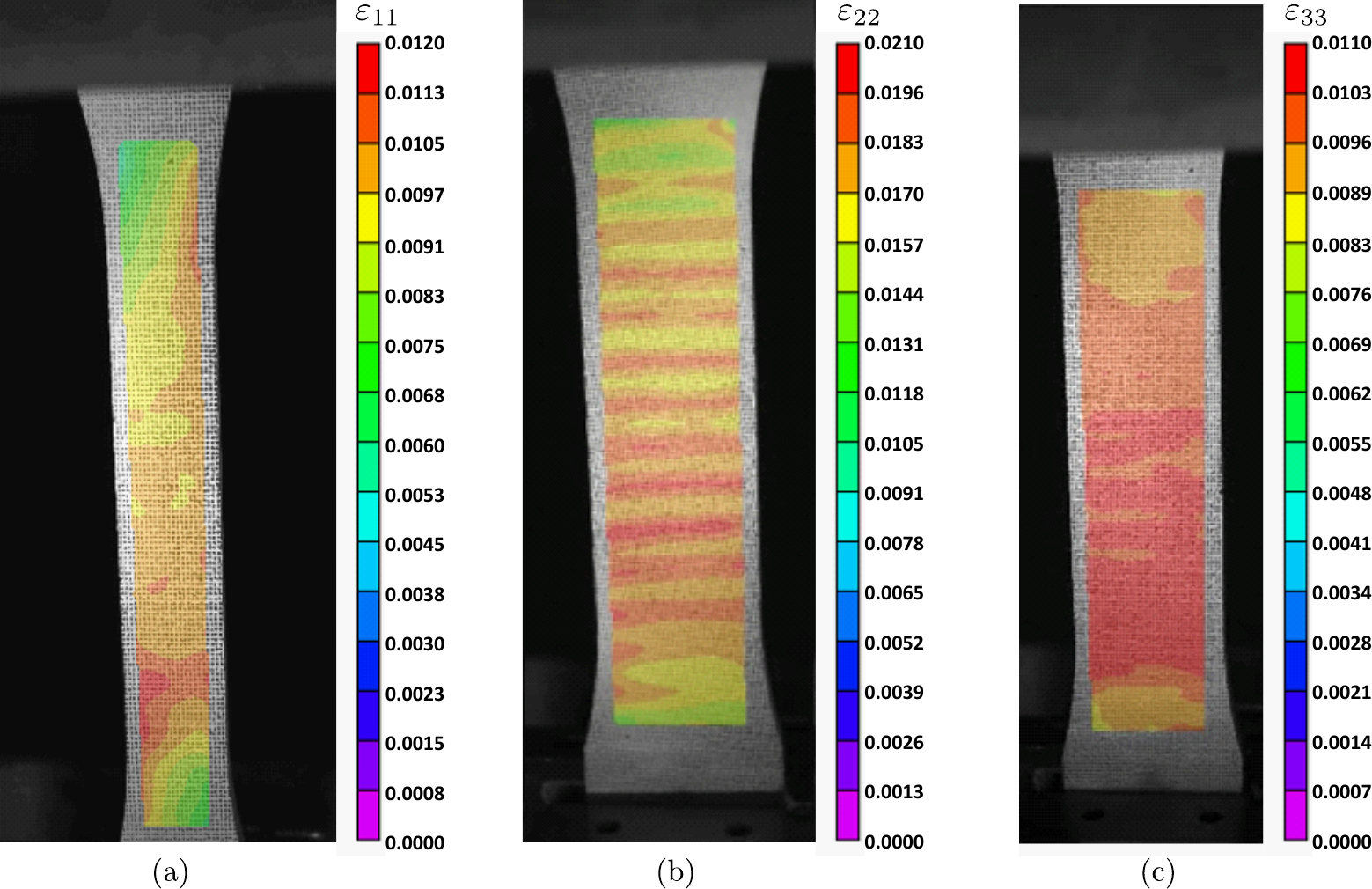} % scale=0.9
  \caption{Strain distribution over a specimen in the (a) 1-direction (b) 2-direction and (c) 3-direction.
   }  \label{fi:DIC}
\end{figure}

\begin{figure}[h]
  \centering
  \includegraphics[width=0.9\linewidth]{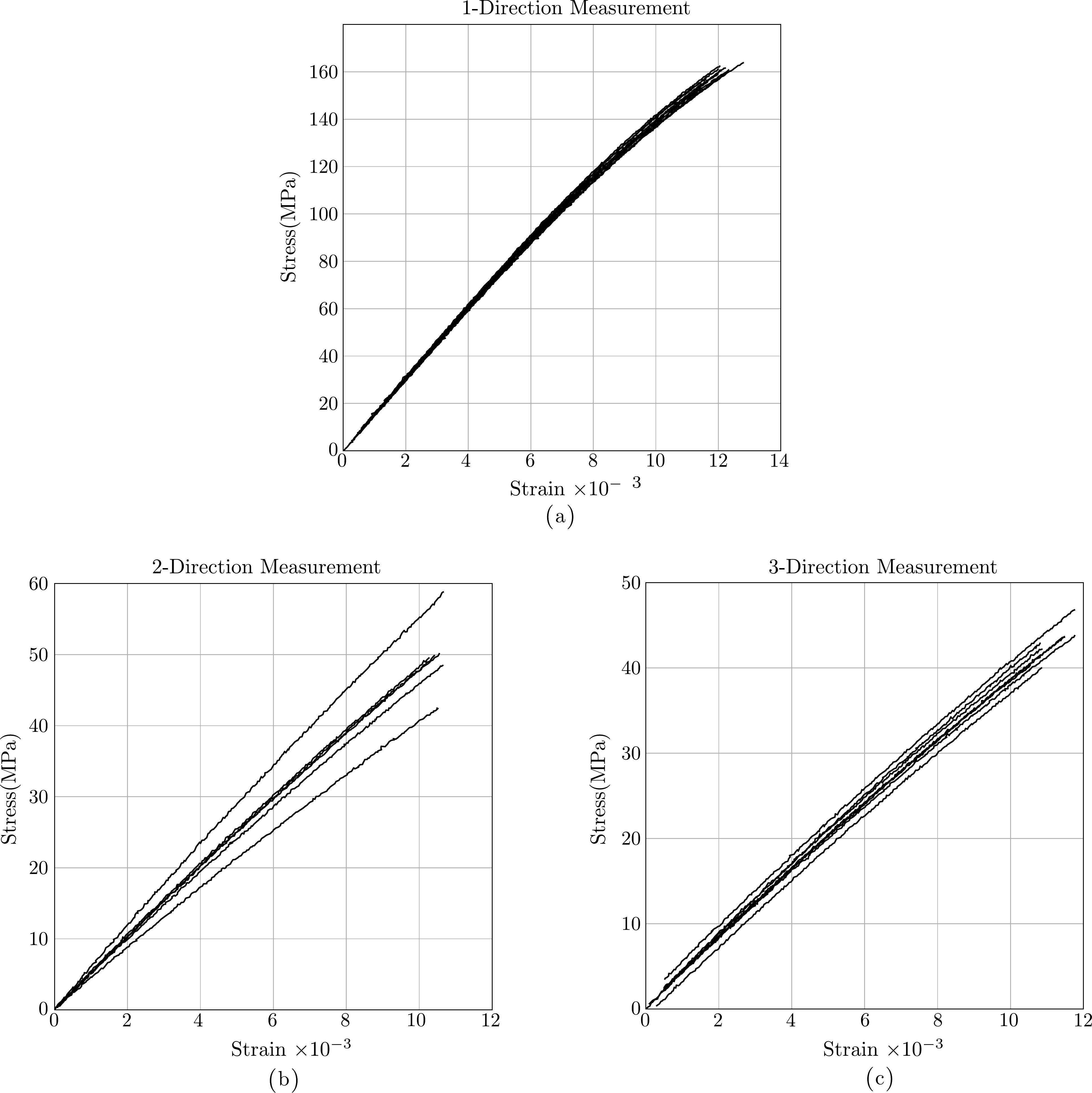} % scale=0.35
  \caption{Stress-strain data obtained from tensile tests in the (a) 1-direction (b) 2-direction and (c) 3-direction.}
  \label{fi:stress_straindata}
\end{figure}

We conducted the tensile tests in accordance with the ASTM D3039 standard~\cite{standard2008standard} on a 98kN MTS load frame and performed the DIC measurement and analysis using a two-camera setup, using a 5-megapixel camera/17 mm lenses combined with the VICSnap and VIC 3D software from Correlated Solutions. 
We illuminated the specimen with a set of lights without overexposing the specimen. 
The specimen was gripped using hydraulic friction wedges and the test was conducted at a rate of 1 mm/min. 
The strain distribution over the entire specimen at one loading instance is shown in \fref{fi:DIC}. 
The colored regions in \fref{fi:DIC} indicate the region of interest selected to measure the strain in each of the experiments. 
We averaged the strain over the entire region of interest to obtain a single value of strain for the corresponding stress applied. 
The stress strain data collected in the three tensile tests are shown in \fref{fi:stress_straindata}. 
We can make some interesting observations from these plots. 
First, we see that the tensile stress-strain data in the 1-direction is consistent without much variation in the stiffness of the composite. 
However, we observe that the 2-direction stress-strain data shows a wide variation of stiffness among the different specimens tested. 
Finally, we observe a systematic bias in the 3-direction stress strain data, i.e., for a particular value of stress, specimens with similar stiffness values show a large variation in the strain values.
Therefore, it is important to model these variations in the Bayesian inference model. 
We explain the mathematical modeling aspects in the following sections.

%-------------------------BAYESIAN INFERENCE -------------------------------%%

\section{Bayesian Inference}
\label{se:BI}
Bayesian inference is the `golden standard'~\cite{srivastava2014dropout} of solving inverse problems. 
The method presented herein, is able to make inferences on the fiber orientation and polymer properties simultaneously while accounting for aleatory (inherent due to the manufacturing process) and epistemic (arising from limited data/information) uncertainty. 
Thus, the parameters to be inferred are considered as random variables whose probability density function (PDF) is to be estimated. 
For simplicity, let $ \tbx $ denote the parameters to be inferred and $\tby$ denote the results of conducted experiments.
Posing a Bayesian inference problem requires three ingredients. First, a PDF over the unknown parameters, $\tbx$, is constructed. 
This PDF, $p(\tbx)$, known as the \textit{prior}, encodes the analyst's prior belief of the unknowns before any data are observed. 
Second, a probabilistic model of the measurement process is constructed. 
This model is known as the \textit{likelihood} and denoted by $p(\tby | \tbx )$. 
The likelihood quantifies the probability of measuring $\tby$ given the parameters  $\tbx$. 
Third, Bayes's theorem is used to get the \textit{posterior} PDF.
The posterior PDF quantifies one's state of knowledge about the unknowns after observing the data.
The posterior PDF is given mathematically by:
\begin{equation}\label{Bayes's_theorem}
    p(\tbx | \tby)  = \frac{p(\tby | \tbx) p(\tbx)} {p(\tby)},
\end{equation}
where $ p(\tby) $ is a normalization constant. 
Typically, the posterior is not analytically available, except for the case of conjugate likelihood-prior pairs~\cite{bishop2006pattern}. 
However, such models fail to represent real world data. Therefore, one has to resort to approximate methods for characterizing the posterior.
One of the most popular approaches is sampling methods, such as the Markov Chain Monte Carlo (MCMC)~\cite{robert1999metropolis}. We use the the No-U-Turn sampler~\cite{hoffman2014no} (as implemented in the Python package pymc3~\cite{salvatier2016probabilistic}) in this work.

The data used for the inference process is the stress strain data and not the average experimental elastic moduli. 
This enables the calibration of experimental noise and systematic errors that could arise during the measurement process.  
Also, the Poisson's ratio of the polymer has a weak correlation to the elastic moduli of the composite. 
Consequently, Bayesian inference results showed wide variation in Poisson's ratio. 
Restricting the Poisson's ratio to physically admissible values with lower variation (using the prior PDF), showed that there was no change in its resulting posterior.
Therefore, we keep the Poisson's ratio of the polymer a constant and choose this value as given by supplier data.
We use a value of 0.365 for the inference. 
~\citet{mohamedou2019bayesian} and~\citet{wu2020bayesian} used a similar method of keeping the Poisson's ratio constant in their work.

\subsection{Likelihood Function}
We now present the likelihood and the priors to characterize the posterior of the fiber orientation and the matrix elastic modulus. 
We assume a linear relationship between stress and strain and the experimental strain values are related to the predicted strain values by:
\begin{equation}\label{strain_relation}
    \epsexp = \epspred + \text{measurement noise},
\end{equation}
where $ \epspred $ is the strain predicted using the micromechanics model and the applied stress on the specimens, and $\epsexp $ is the experimentally measured strain values. 
By limiting the strain data to the elastic regime, we can write:
\begin{equation}\label{stress-strain-realtionship}
    \epspred = \frac{\sigmaexp}{E_c(\bx)}.
\end{equation}
Here, $E_c (\bx)$ is the composite elastic stiffness predicted by the micromechanics model as a function of the parameters to be inferred, $\bx$.  
The variable $\sigmaexp$ represents the stress applied on the specimen and we assume it to be noise free in this investigation. 
Assuming normal measurement noise with standard deviation $\tauexp$ (which is to be determined) and independent measurements, we can write the likelihood as:
\begin{equation} \label{likelihood}
\begin{split}
p(\epsexp | \sigmaexp, \bx, \tauexp, b_d)\ & = \prod_{i=1}^{K} \prod_{j=1}^{N} p(\varepsilon_{ij} | \bx_i, \sigma_{ij}, \tauexp, b_d )\\
 & = \prod_{i=1}^{K} \prod_{j=1}^{N} \mathcal{N} (\varepsilon_{ij} | \frac{\sigma_{ij}}{E_c (\bx_i) } +b_d, \tauexp^2).
\end{split}
\end{equation}

Since the linear stress-strain relation is used to model the likelihood, only experimental strain data below $5 \times 10^{-3} $ were considered in the inference process (stress strain data is shown in \fref{fi:stress_straindata}). 
\qref{likelihood} represents the joint probability of observing the strain values given the values of the unknowns (values to be inferred). 
The double product spans $N$ data points in each of the $K$ samples tested. 
Note, that $K$ represents the total number of specimens tested across the three tensile tests conducted. 
The extra parameter, $b_d$, accounts for any systematic errors introduced in the strain measurements while positioning and clamping the specimen in the load frame. The subscript $d$ stands for `direction'. For example, $b_1$, represents the bias variable that accounts for the systematic error for all the specimens tested in the 1-direction. 

\subsection{Priors and Hyperpriors}
Having defined the likelihood, we now define the priors. 
Priors enforce constraints to be placed on the parameters to be inferred and the analyst's belief of the parameter values before any data is observed. 
We use a hierarchical Bayesian modeling approach in this study where we write the model in multiple levels that capture population behavior of the estimated parameters across different samples. 
Therefore, prior distributions are parametrized by random variables, which themselves have a prior distribution, called the hyperpriors. 
\citet{gelman1995bayesian} gives a detailed description of this approach. 

We place a normal prior with mean $\mu_{M}$ and standard deviation $\tau_M$ on the matrix modulus, $E_i$, as: 
\begin{equation} \label{E-prior}
 E_i \sim \mathcal{N} (\mu_M, \tau_M^{2}).
\end{equation}
In this case, the distribution on $E_i$ represents the aleatory uncertainty. 
Here, we place a prior on the matrix moduli value for each of the $K$ specimens tested (and hence denoted by $E_i$).
We reparametrize the orientation values, $a_{11}$ and $a_{22}$, using $\theta_{0}$ and $\phi_{0}$ as:
\begin{equation}\label{a11-prior}
a_{11i} = \sin^2 (\theta_{0i}) \cos^2 (\phi_{0i}),
\end{equation}
and
\begin{equation}\label{a22-prior}
a_{22i} =\sin^2 (\theta_{0i}) \sin^2 (\phi_{0i}).
\end{equation}
% \begin{gather}
% % \theta_{0i} \sim \mathrm{LogNormal} ( \mu_{\theta_{0}}, \tau_{ \theta_{0} }), \\
% % \phi_{0i} \sim \mathrm{LogNormal} ( \mu_{\phi_{0}}, \tau_{ \phi_{0} }), \\
% a_{11i} = sin^2 (\theta_{0i}) cos^2 (\phi_{0i}),\\
% \label{a22-prior}
% \begin{split}
%     a_{22i} &= sin^2 (\theta_{0i}) sin^2 (\phi_{0i}),\\
%      & \mathrm{for} \ i = 1\dots K.
% \end{split}
% \end{gather}
% \ibsays{Give this equation the respect it requires. It is not a figure.
% Write it in a way we can read it.
% Also, the $a_{11}$ and the $a_{22}$ should depend on $i$, right?
% }
There are two primary reasons for this reparameterization. First, this enforces the physical constraint enforced by the statistical description of the orientation tensor: $a_{11} + a_{22} \leq 1$. Second, it helps in exploring the posterior more efficiently by projecting the orientation values into an unconstrained space. 
Since $a_{11}$ and $a_{22}$ are expected to be highly correlated, their joint posterior would be extremely hard to explore otherwise.
The priors on $\theta_{0}$ and $\phi_{0}$ are chosen as:
\begin{gather} 
\theta_{0i} \sim \mathrm{LogNormal} ( \mu_{\theta_{0}}, \tau_{ \theta_{0} }), \\
\phi_{0i} \sim \mathrm{LogNormal} ( \mu_{\phi_{0}}, \tau_{ \phi_{0} }).
\end{gather}
respectively. Next, we place a normal prior on the measurement noise standard deviation,  $\tauexp$ , given by: 
\begin{equation}\label{noise_prior}
    \tauexp \sim \mathrm{Exponential} ( \lambda_e )
\end{equation}
Finally, we choose the priors on the bias variables that account for the systematic as:
\begin{equation}\label{bias_prior}
    \begin{split}
        & b_d \sim \mathcal{N} (0, \lambda_d) \\ 
        & \mathrm{for} \ d= 1,2,3.
    \end{split}
\end{equation}
Recalling that there are three tensile tests conducted (i.e., in the printing direction, transverse to the printing direction in the printing plane, and the stacking direction), we use three bias random variables in the Bayesian model. 
Each $b_d$ for $ d = 1, 2, 3$ corresponds to the bias variable in the 1-direction, 2-direction, and 3-direction respectively. 

We use hyperiors only for the material parameters being inferred with the motivation that we are interested in the population behavior of these parameters.
The hyperpriors we use for the polymer modulus, $E_i$, are:
\begin{gather}
\mu_M \sim \mathcal{N} (0, 0.8), \\
\tau_M \sim \mathrm{HalfNormal}(1).
\end{gather}
Next, hyperpriors placed on $\theta_0$ are:
\begin{gather}
    \mu_{\theta_0} \sim \mathcal{N} (0, 4),\\
    \tau_{\theta_0} \sim \mathrm{HalfNormal}(2).
\end{gather}
Finally, hyperpriors we use for $\phi_0$ are:
\begin{gather}
\mu_{\phi_0} \sim \mathcal{N} (0, 4),\\
\tau_{\phi_0} \sim \mathrm{HalfNormal}(2).
\end{gather}

\begin{figure}[ht]
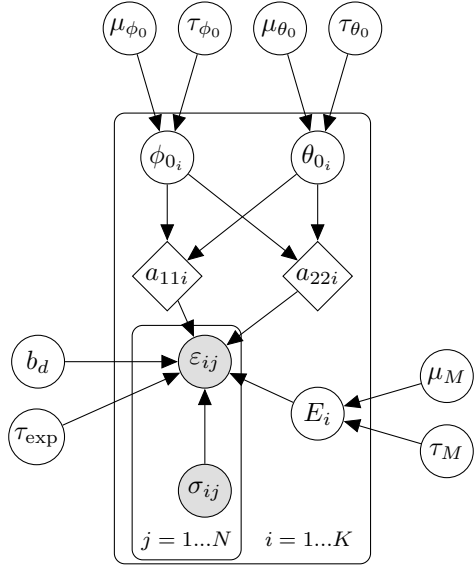

  \centering
  \tikz{ %
    
    \node[obs, yshift = 2cm] (strain) {$\varepsilon_{ij}$} ; %
    \node[obs, below=of strain] (sigma) {$\sigma_{ij}$} ;
    \node[latent, left=of strain, xshift = -0.5 cm, yshift = -1 cm] (noise) {$\tauexp$};
    \node[latent, left=of strain, xshift = -0.5 cm, yshift = 0 cm] (bias) {$b_d$};
    \node[latent, above= of strain, xshift = -0.5 cm, yshift = 1 cm] (phi0) {$\phi_{0_{i}}$};
    \node[latent, above= of strain, xshift = 1.5 cm, yshift = 1 cm] (theta0) {$\theta_{0_{i}}$};
   
    \node[det, below=of phi0, yshift = 0.25 cm] (a11) {$a_{11i}$};
    
    \node[det, below=of theta0, yshift = 0.25 cm] (a22) {$a_{22i}$};
    
    \node[latent, below= of a22,xshift = 0 cm, yshift = 0 cm] (matrix_modulus) {$E_{i}$};
    \node[latent, above =  of phi0, xshift = -0.5 cm](phi_mean) {$\mu_{\phi_{0}}$};
    \node[latent, above =  of phi0, xshift = 0.5 cm](phi_sd) {$\tau_{\phi_{0}}$};
    
    \node[latent, above =  of theta0, xshift = -0.5 cm](theta_mean) {$\mu_{\theta_{0}}$};
    \node[latent, above =  of theta0, xshift = 0.5 cm](theta_sd) {$\tau_{\theta_{0}}$};
    
    \node[latent, right = of matrix_modulus, yshift = 0.5cm] (matrix_modulus_mean) {$\mu_M$};
    \node[latent, right = of matrix_modulus, yshift = -0.5cm] (matrix_modulus_sd) {$\tau_M$};
    \plate  [inner sep=0.225cm]  {plate2}{(sigma) (strain) (a11) (a22) (theta0) (matrix_modulus) (phi0)} {$i=1...K$}; %
    %\plate[inner sep=0.25cm, xshift=-0.12cm, yshift=0.12cm] {plate1} {(sigma) (strain)} {$j=1...N$}; 
    \plate {plate1} {(sigma) (strain)} {$j=1...N$};

    \edge {sigma} {strain} ; %
    \edge {a11} {strain} ;
    \edge {a22} {strain} ;
    \edge {matrix_modulus} {strain} ;
    \edge {theta0} {a11, a22};
    \edge {phi0} {a11, a22};
    
    \edge {bias} {strain};
    \edge {noise} {strain};
    
    \edge{phi_mean}{phi0};
    \edge{phi_sd}{phi0};
    \edge{theta_mean}{theta0};
    \edge{theta_sd}{theta0};
    \edge{matrix_modulus_mean}{matrix_modulus};
    \edge{matrix_modulus_sd}{matrix_modulus};

    }

    \caption{Probabilistic graph illustrating the Bayesian model.}
    \label{fi:prob}
\end{figure}

The probabilistic statements made above can be graphically illustrated using a directed acyclic graph (DAG) that makes use of the plate notation~\cite{bishop2006pattern}.
This is shown in \fref{fi:prob}. 
In a graphical representation of a probabilistic model, the nodes which are denoted by circles, represent random variables and their conditional independence is represented using the edges. 
In \fref{fi:prob} the the shaded circles indicate variables that are observed, i.e., variables which are experimentally measured.
Here the strain values, $\varepsilon_{ij}$, and stress values, $\sigma_{ij}$, are experimentally measured and therefore appear in the likelihood. 
The un-shaded nodes are unobserved and need to inferred in the model. 
The plate (bounding box) represents the joint distribution of random variables. Therefore, here, the index $j$ spans strain data for a particular specimen, $i$, which was tested. 

Note that we perform the inference of the parameters on a standardized space. 
We replace the the micromechanics model with a Gaussian Process surrogate for computational feasibility. 
Since we train the surrogate using standardized inputs, we use the same standardized space for the inference.
We explain the surrogate training and testing procedures in the following section.

%-----------------------------------------------------------------------------------%%

%-------------------------GAUSSIAN PROCESS SURROGATE -------------------------------%%
\section{Gaussian Process Regression Based Surrogate}
\label{se:GPR}

Characterizing the posterior of the parameters using MCMC methods parameters require samples in the order of $10^4 - 10^6$ to ensure the convergence of the Markov chain to its stationary distribution~\cite{betancourt2017conceptual}.
This is computationally infeasible even when the micromechanics model takes approximately 1.0 s to evaluate. 
To evaluate the micromechanics model discussed previously, it takes approximately 0.5s. 
Therefore, we use a probabilistic surrogate to replace the micromechanics model. 
In this study, we use the mean predictions of Gaussian Process, ($\GP$) conditioned on a finite number of model evaluations as the micromechanics surrogate.

Formally, a GP is a collection of random variables from a function space such that any finite subset of these random variables have a joint Gaussian distribution~\cite{bishop2006pattern}. 
Consider a dataset, $\mathcal{D}  := (\bx^m,g^m)$, where $m=1\dots M$. 
To clarify, $\mathcal{D}$ is dataset of $M$- input-output pairs. Here, $\bx = (E,\nu, a_{11},a_{22})$ and $\mathbf{g}=(E_1,E_2,E_3)$. 
The variable $\nu$ represents the Poisson's ratio of the polymer. Note that, $\mathbf{g}$ has a dimension of 3 and we use the notation $g_r =f_r (\bx)$ to refer to each of the composite stiffness values. 
We model each $f_r(\bx)$ using a $\GP$ defined by a mean function and a covariance function:
\begin{equation}\label{GP_mean_cov}
 f_r(\bx) | \boldsymbol{\theta} \sim \GP (m(\bx), k(\bx, \bx'; \boldsymbol{\theta})),
\end{equation}
where $\boldsymbol{\theta}$ are the hyperparameters of the covariance function. Our prior beliefs about the function (length-scales, variance, smoothness) are encoded using the mean and covariance functions. In this study, we choose a zero-mean function, which implies that we do not enforce any prior belief on the function mean value, and the squared exponential covariance function defined as:
\begin{equation}\label{Covariance_GP}
    k(\bx^i, \bx^j; \boldsymbol{\theta}) = s^2 \exp \left (-\frac{1}{2} (\bx^i - \bx^j)^T \Lambda^{-1} (\bx^i - \bx^j) \right).
\end{equation}
The squared exponential  covariance function encodes the belief that the model output is infinitely differentiable with respect to the input parameters. The hyperparameters in this model are $ \boldsymbol{\theta} = (s, l_1, l_2, l_3, l_4)$. 
The parameter, $s$, is called the signal strength and the parameters  $ l_1, l_2, l_3,$ and $l_4$ are called the characteristic length scales. 
The characteristic length scale can be informally thought of the distance that needs to be moved in parameter space before a significant change in the function is observed.
As a result, a larger length scale in a particular parameter space implies that the function output is less sensitive to changes of that particular parameter. 
The lengthscales are incorporated into the covariance function by $ \Lambda = \mathrm{diag} (l_1^2, l_2^2, l_3^2, l_4^2)$. 

Denoting the given input data set as $ \mathcal{D}_X$, the prior $\GP$ (using the zero mean function and covariance function) on the function response is:
\begin{equation}\label{prior_GP}
    \textbf{f}_\textbf{r} | \mathcal{D}_X, \boldsymbol{\theta} \sim \mathcal{N} (\textbf{0}, \textbf{K})
\end{equation}
where, $\textbf{K}$ represents the covariance matrix with an entry at $(i,j)$ defined as $k(\bx^i, \bx^j; \boldsymbol{\theta})$. 
Note that, \qref{prior_GP} represents a joint probability of the function responses conditioned on the inputs before observing any data.

A Gaussian noise can be added to each function output to obtain noisy outputs (since datasets are often contaminated with noise) defined as:
\begin{equation}\label{outputs_noisy}
    t_r^{(m)}  = f_r(\bx^{(m)}) + \epsilon_r^{m},
\end{equation}
where $ \ m = 1 \dots M $. The variable $\epsilon_r$ represents a Gaussian noise vector with zero mean and standard deviation of $\tau_{\GP,r} $, i.e., $\epsilon_r \sim \mathcal{N} (0, \tau_{\GP,r}^2)$. 
Here, we use the $\GP$ to build a surrogate for simulation outputs, which implies that the use of the noise vector is not necessary. 
However, we include it in the training since it aids numerical conditioning during the factorization of the covariance matrix~\cite{ababou1994condition}.

Introducing a set of test inputs, denoted by $\mathcal{D}_{X^*}$, and invoking that any finite subset of random variables in a $\GP$ are jointly Gaussian, the joint distribution of the observed noisy target and the function evaluations at the target inputs can be written as:
\begin{equation}\label{joint_posterior}
    \begin{bmatrix}  \textbf{t}_r \\ \textbf{f}^*_r  \end{bmatrix}
 \sim \mathcal{N} \left ( \textbf{0},  \begin{bmatrix}  \textbf{K}(\mathcal{D}_X, \mathcal{D}_X) + \tau_{\GP,r}^2 \textbf{I} &   \textbf{K}(\mathcal{D}_X, \mathcal{D}_{X^*})  \\ \textbf{K}(\mathcal{D}_{X^*}, \mathcal{D}_{X}) &  \textbf{K}(\mathcal{D}_{X^*}, \mathcal{D}_{X^*}) \end{bmatrix}\right), 
\end{equation}
where $\textbf{f}^*_r$ represents the vector of function outputs evaluated at the inputs, $\mathcal{D}_{X^*}$, and $\textbf{K}(\mathcal{D}_X, \mathcal{D}_{X^*}) $ denotes the $M \times M_*$ (where $M$ denotes the size of the training set and $M_*$ denotes the size of the test set)matrix of the covariance values evaluated at all pairs of the training and test data points. 
Following~\cite{rasmussen2003gaussian}, Baye's rule is used to evaluate the posterior distribution of $\textbf{f}^*_r$ from which predictive mean and variance at a single test point, $\bx^*$ $\in$ $\mathcal{D}_{X^*}$ can be written as:
% \begin{equation}\label{posterior_of_f}
%     \textbf{f}_r^* | \mathcal{D}_{X}, \textbf{t}_r, \mathcal{D}_{X^*}, \boldsymbol{\theta} \sim \mathcal{N} \left ( \Bar{\textbf{f}^*}_r, \mathrm{cov}(\textbf{f}^*_r)  \right),
% \end{equation}
% \ibsays{This is a weird thing to write. Do not write it. This will be shortened anyway. So I am not spending time here.}
% where,

% \begin{gather}
% \label{posterior_predictive_mean}
% \Bar{\textbf{f}^*}_r  := \mathbb{E} [\textbf{f}^* | \mathcal{D}_{X}, \textbf{t}_r, \mathcal{D}_{X^*}, \tau_{\GP,r} ] = \textbf{K}(\mathcal{D}_{X^*}, \mathcal{D}_{X}) [ \textbf{K}(\mathcal{D}_X, \mathcal{D}_X) + \tau_{\GP,r}^2 \textbf{I} ]^{-1} \textbf{t}_r, \\
% \label{psoterior_predictive_coveriance}
% \mathrm{cov} (\textbf{f}^*_r)  =  \textbf{K}(\mathcal{D}_{X^*}, \mathcal{D}_{X^*})  - \textbf{K}(\mathcal{D}_{X^*}, \mathcal{D}_{X})  [ \textbf{K}(\mathcal{D}_X, \mathcal{D}_X) + \tau_{\GP,r}^2 \textbf{I}]^{-1}  \textbf{K}(\mathcal{D}_X, \mathcal{D}_{X^*}).
% \end{gather}

% Note that equations \ref{posterior_predictive_mean} and \ref{psoterior_predictive_coveriance} represents the mean vector and the covariance matrix respectively of the posterior of $\textbf{f}_r^*$. 
\begin{equation}\label{point_predictive_mean}
    \mu_{f_r} (\bx^*; \boldsymbol{\theta}, \tau_{\GP,r} ) = \textbf{k}(\bx^*, \mathcal{D}_X ; \boldsymbol{\theta}) ( \textbf{K}(\mathcal{D}_X, \mathcal{D}_X) + \tau_{\GP,r}^2 \textbf{I})^{-1} \textbf{t}_r,
\end{equation}

\noindent
and
\begin{equation}\label{point_predictive_variance}
    \tau_{f_r} ^2 (\bx^*; \boldsymbol{\theta}, \tau_{\GP,r} )  = k(\bx^*, \bx^* ; \boldsymbol{\theta})  -  \textbf{k} (\bx^*, \mathcal{D}_X ; \boldsymbol{\theta} ) \textbf{K}(\mathcal{D}_X, \mathcal{D}_X) + \tau_{\GP,r}^2 \textbf{I})^{-1}   \textbf{k} ( \mathcal{D}_X, \bx^*; \boldsymbol{\theta} ) 
\end{equation}
respectively.

The mathematical formulation of the $\GP$ has now been setup.
However, we still need to find the values of the hyperparameters, $\boldsymbol{\theta}$ and $\tau_{\GP,r} $.
This step of finding $\boldsymbol{\theta}$ is called the Gaussian Process Regression (GPR).
We need to choose the hyperparameters such that we maximize the likelihood of the output data in the training set. 
For numerical stability (in cases where likelihood is extremely small or large), log-likelihood is preferred. 
Then, the optimal values of the hyperparameters, $\theta^{\prime}$ and $\tau_{\GP, r}^{\prime}$ are found by solving an optimization problem~\cite{lee2019predicting} defined as:
\begin{equation}\label{max_log_likelihood_GP_OPT}
    \boldsymbol{\theta}^\prime, \tau_{\GP,r}^\prime =  \underset{\theta,  \tau_{\GP,r}}{\operatorname{argmax}} \ \mathcal{L}_{\GP}( \boldsymbol{\theta}, \tau_{\GP,r} ; \mathcal{D}_X, \textbf{t}_r),
\end{equation}
where $\mathcal{L}_{\GP}$ denotes the likelihood of the training data. 

\begin{figure}[ht]
  \centering
  \includegraphics[width=0.85\linewidth]{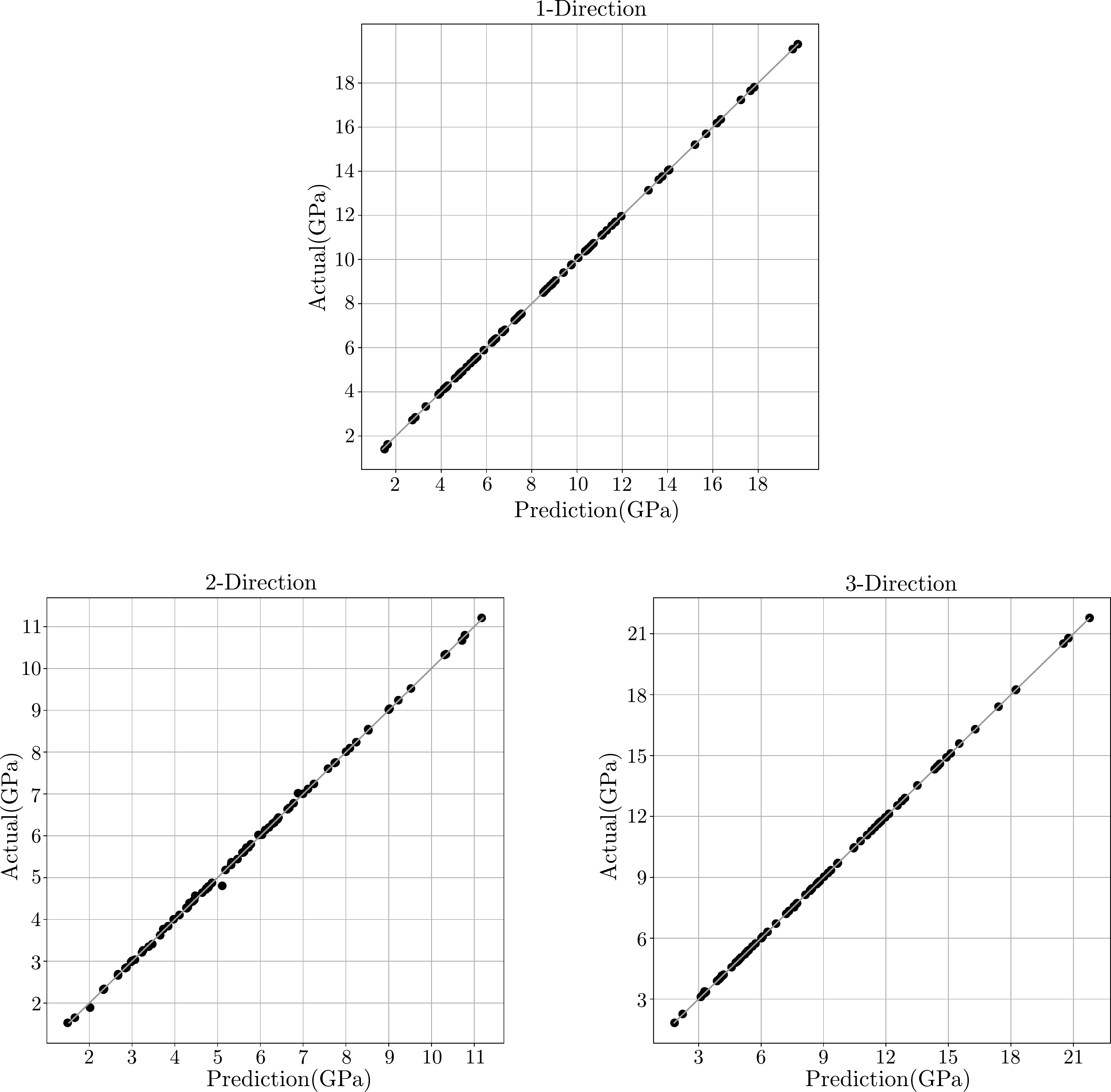} % scale=0.32
  \caption{Parity plots for comparison of predicted stiffness vs actual stiffness in the 1-direction (printing direction), 2- direction (transverse to printing direction), and 3-direction (stacking direction) }
  \label{fi:Parity_GP}
\end{figure}

We implement the GPR using the python package pymc3~\cite{salvatier2016probabilistic} and use the Broyden Fletcher Goldfarb Shanno (BFGS) algorithm for the hyperparameter optimization.
Maximizing the log-marginal likelihood is a non-convex problem~\cite{rasmussen2003gaussian}. 
To ensure we obtain the best possible optimal value of the hyperparameters, we restart the algorithm at a total of 5 different starting points. 
We choose the hyperparameters that result in the highest log-marginal likelihood. 
We generate a training data set of size M = 288 by sampling the inputs using the Latin Hypercube design~\cite{moon2011algorithms} and running the micromechanics model for the corresponding inputs.
Note that, this data set is created by initially sampling a dataset of size 350 data points and rejecting the data points that violate the physical constraint : $a_{11} + a_{22} \le 1$ enforced by the statistical description of the orientation tensor.
To evaluate the predictions of the surrogate, we create a test data set of size 81 points using the Latin Hypercube Design. 
The prediction (from the predictive mean of the GP) vs the actual stiffness values for the test data set is shown in \fref{fi:Parity_GP}. 
The Mean Squared Error (MSE) on the test data for each of the stiffness prediction are given in table \ref{table:MSE_GP}.
From \fref{fi:Parity_GP} and table \ref{table:MSE_GP} we see that the GPR surrogate predicts the three composite stiffness well.

\begin{table}[]
\caption{Mean Squared Error for GP surrogates} % title of Table
\centering % used for centering table
\begin{tabular}{| c || c |} % centered columns (4 columns)
\hline\hline %inserts double horizontal lines
Surrogate Model & MSE \\ [0.5ex] % inserts table
%heading
\hline % inserts single horizontal line
$E_1$ Surrogate & 0.0002\\ % inserting body of the table
$E_2$ Surrogate & 0.0021 \\
$E_3$ Surrogate & 0.0003 \\
\hline %inserts single line
\end{tabular}
\label{table:MSE_GP} % is used to refer this table in the text
\end{table}

%-----------------------------------------------------------------------------------%%
\section{Results and Discussion}
\label{se:res_and_dis}

\begin{figure}[ht]
  \centering
  \includegraphics[width=0.92\linewidth]{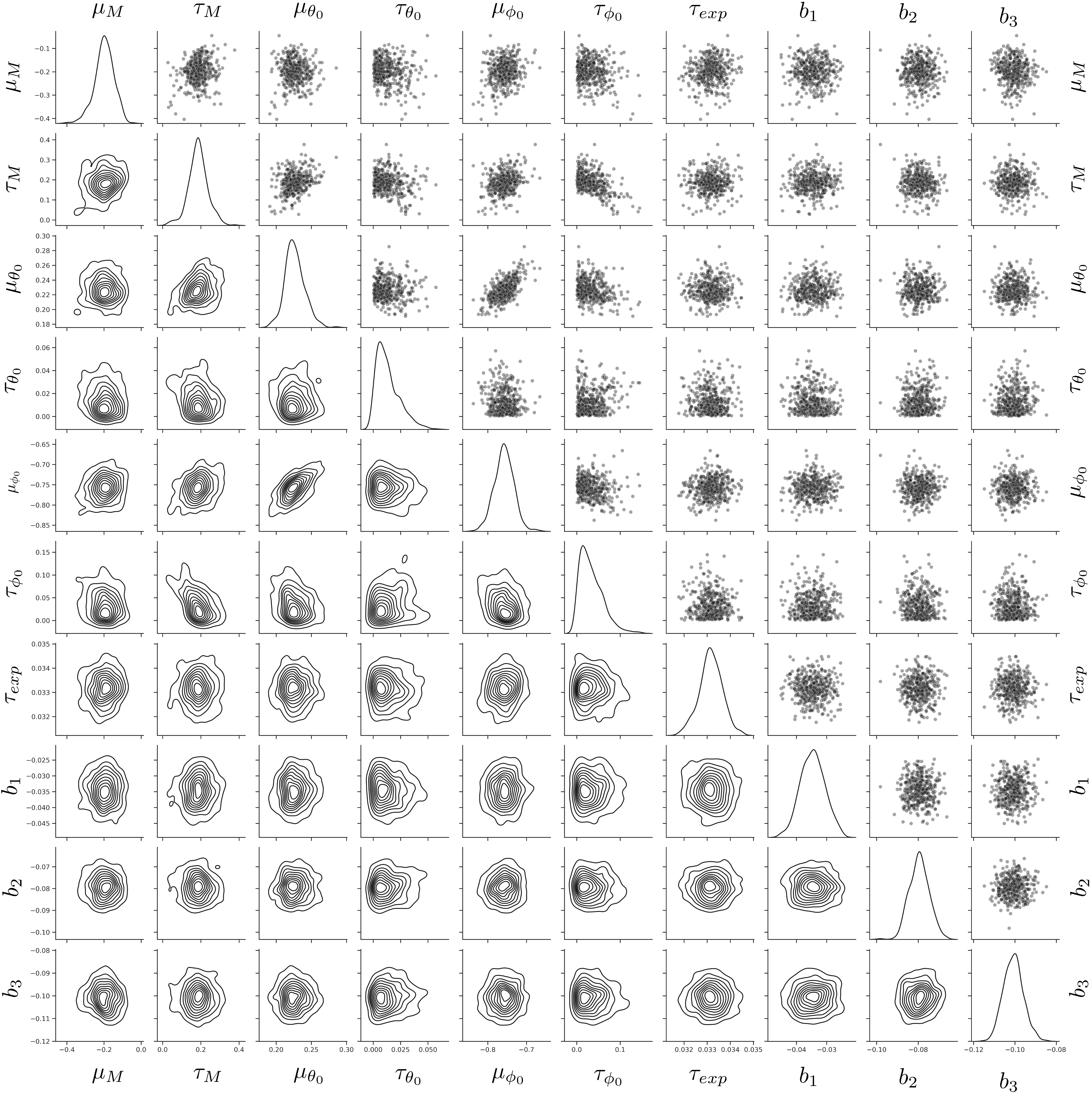} % scale=0.22
  \caption{Identified marginal and joint distributions of the population parameters, experimental noise, and bias variables}
  \label{fi:Pair_plot}
\end{figure}

Once we train the $\GP$ surrogate, we can now perform the Bayesian inference sampling. 
To ensure the convergence of the Markov chain to the stationary distribution, we draw a total of 175,000 samples. 
We discard the first 40,000 samples to remove transient behavior and from the samples that remain, we keep one out of every 40 to reduce the chain autocorrelation.
The results of the Bayesian inference are samples from the probability distribution functions (PDF) of the inferred properties.
The marginal and the joint distributions of the population parameters, experimental noise, and bias variables are shown in \fref{fi:Pair_plot}. 
We study the joint distributions of these variables since they control the data generation process. 
Once we know the posterior of these variables, we can generate posterior samples of the strain data using the Bayesian model discussed in the earlier section.
We can find the marginalized distributions of the matrix modulus, and the orientation values - $a_{11}$, $a_{22}$ by integrating out the population parameters using the sum rule. We discuss this shortly.
From \fref{fi:Pair_plot} we see that the joint distributions of these variables do not show any strong correlations, indicating our Bayesian model has not been over parameterized.

\begin{figure}[h]
  \centering
  \includegraphics[width=0.93\linewidth]{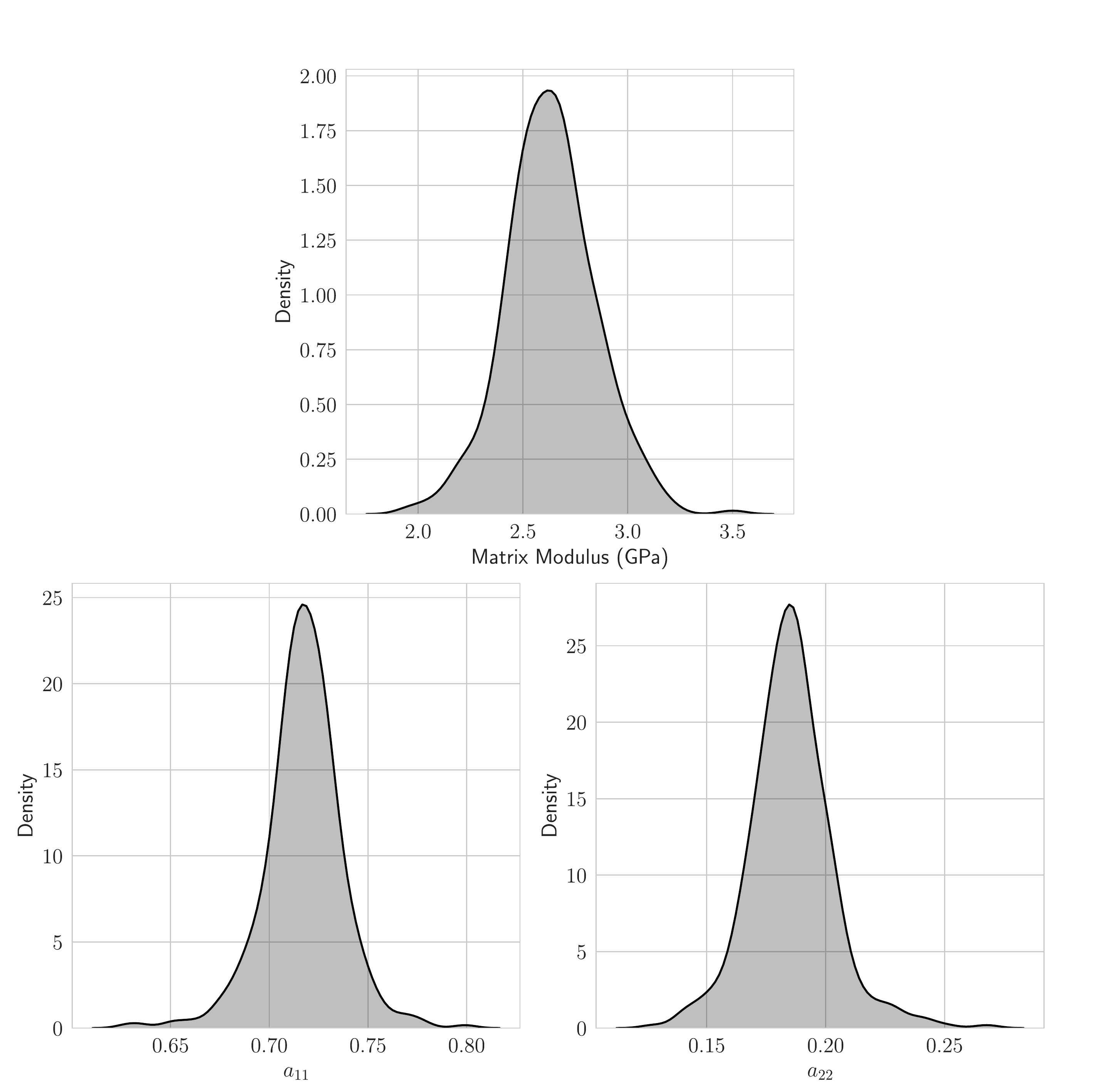} % scale=0.52
  \caption{Inferred  PDFs of matrix modulus ($E$), $a_{11}$, and $a_{22}$ }
  \label{fi:Marginalized_Properties}
\end{figure}

To obtain the PDF of the material properties, we need to perform a marginalization step which integrates out the population parameters.
For brevity, we demonstrate the mathematical setting to characterize the marginalized posterior of the matrix modulus, $E$. 
Let $\textbf{X}$ denote the data set used for the inference (collected from experiments). Recalling that the population parameters for $E$ are  $\mu_M$ and $\tau_M$ (described earlier in \fref{fi:prob}), the marginalized posterior for $E$ can be written using the sum rule as: 
\begin{equation}\label{Marginalization_of_E}
    p(E | \textbf{X}) = \int_{R_{\mu_M}, R_{\tau_M}} p(E |\mu_M, \tau_M ) p(\mu_M, \tau_M | \textbf{X}) d \mu_M d \tau_M,
\end{equation}
where $p(E |\mu_M, \tau_M )$ is evaluated using the prior on $E$ and $p(\mu_M, \tau_M | \textbf{X})$ is obtained from the posterior. 
Note that $p(\mu_M, \tau_M | \textbf{X})$ is the marginalized posterior for the population parameters which is directly obtained as a results of the MCMC sampling. $R_{\mu_M}$ and $R_{\tau_M}$ indicates the support of $\mu_M$ and $ \tau_M$ respectively. 
Following a similar method, we can marginalize the orientation values. 
Since the results from the MCMC evaluations are samples from the posterior, we plot the PDF for the material properties via a kernel density plot using the Gaussian kernel.  
This is shown in \fref{fi:Marginalized_Properties}.
The mean values of the distributions are shown in table \ref{table:Mean_Bayesian_Values}.
It is important to note here that the distributions inferred (and shown in \fref{fi:Marginalized_Properties}) are ``effective distributions" that best matches the strain data. 
As seen earlier in \fref{fi:DIC}, there is variation of strain within the length of each specimen. 
However, we record only the average strain value across the region of interest. 
As a consequence, the Bayesian model infers parameter distribution across the different specimens tested, rather than the distribution within a particular specimen. 
Since we extracted the specimens from different regions of an additively manufactured plate/wall, there exists variation in the inferred properties. 
Therefore, the variation of these parameters across different regions of the manufactured plate/wall manifest themselves as the primary source of aleatory uncertainty. 
To capture more variation in the inferred parameters, we need to define smaller regions of interest across a specimen for the strain measurement. 
This, however, raises the question of choosing the size of the smallest region of interest that captures the true uncertainty of the parameters and is beyond the scope of this current study.

%The focus of this work is to identify the composite elasticity tensor with few experimental tests. 

\begin{table}[h]
\caption{Mean value of the inferred properties} % title of Table
\centering % used for centering table
\begin{tabular}{| c || c |} % centered columns (4 columns)
\hline\hline %inserts double horizontal lines
Parameter & Inferred Values \\ [0.5ex] % inserts table
%heading
\hline % inserts single horizontal line
Matrix Modulus, $E (GPa)$ & 2.63\\ % inserting body of the table
$a_{11}$ & 0.72 \\
$a_{22}$ & 0.18 \\
$a_{33} = 1- a_{22} - a_{11}$ & 0.10\\
\hline %inserts single line
\end{tabular}
\label{table:Mean_Bayesian_Values} % is used to refer this table in the text
\end{table}

\begin{figure}[h]
  \centering
  \includegraphics[width=0.75\linewidth]{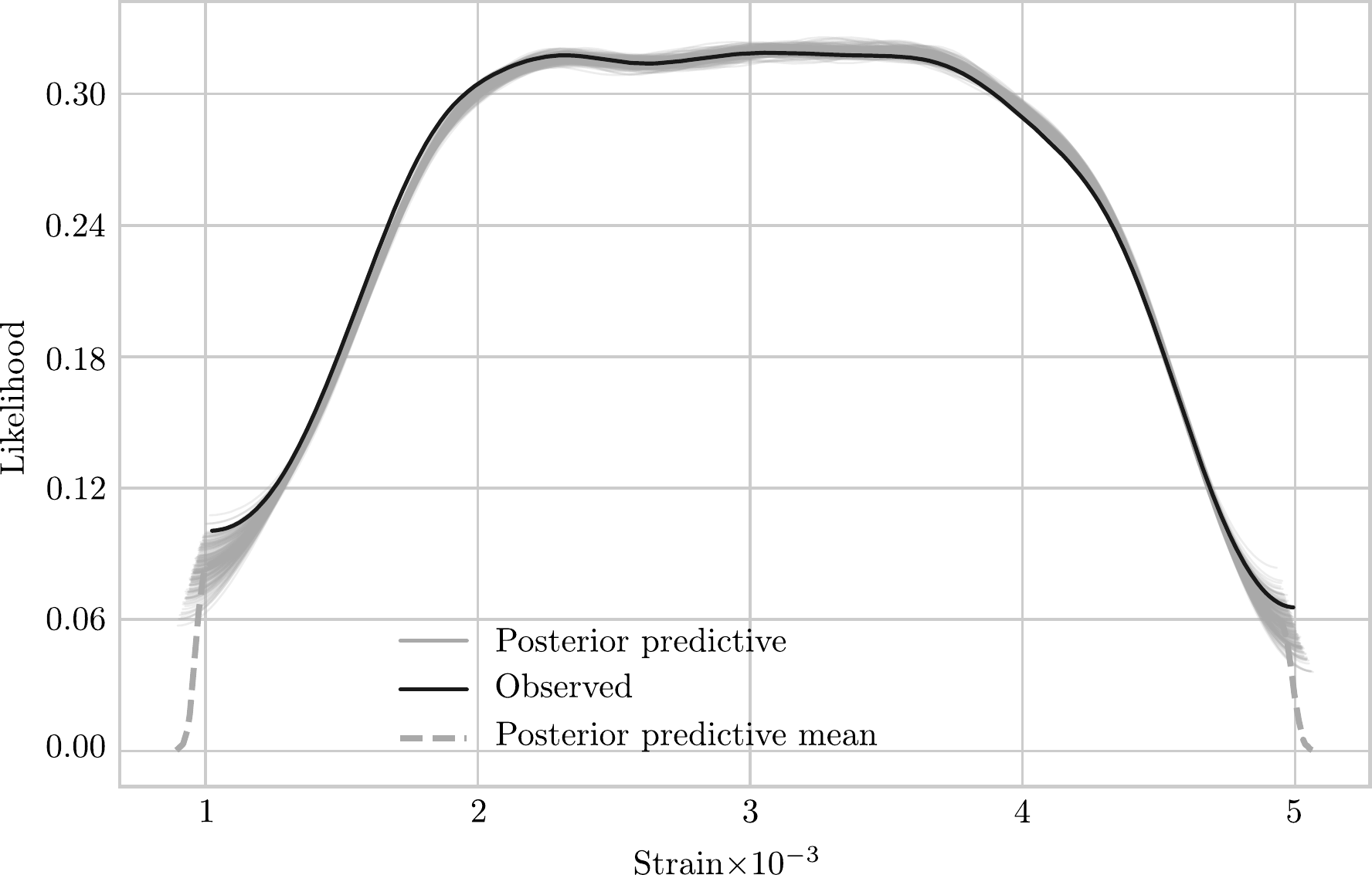} % scale=0.60
  \caption{Posterior predictive distribution of strain values vs observed strain values.
  }
  \label{fi:Post_pred_strain}
\end{figure}

Next, we need to conduct a set of tests to validate the inferred parameters.
To evaluate the inferred distributions, we perform a posterior predictive check, i.e., we check if the inferred parameters can reproduce the experimental strain data. 
We can evaluate the posterior predictive distribution again using the sum rule as:
\begin{equation}\label{posterior_predictive}
    p({\varepsilon} | \boldsymbol{\sigma}, \mathcal{D}, \textbf{X}, \boldsymbol{\theta} )
                        = \int_{R_{\Theta}}  p({\varepsilon} | \boldsymbol{\sigma}, \mathcal{D}, \boldsymbol{\theta}, \Theta)
                                p(\Theta | \textbf{X}, \boldsymbol{\theta}) d \Theta,
\end{equation}
where $\mathcal{D}$ is the dataset used for GPR,  $\boldsymbol{\sigma}$ is the experimental stress data  used for inference, $\boldsymbol{\theta}$ are the hyperparameters for the GPR model, and $\Theta$ are the latent variables in the probabilistic inference model (the un-shaded nodes in  \fref{fi:prob} are the latent variables in the model). 
The posterior predictive vs observed strain value plot can be seen in \fref{fi:Post_pred_strain}.
From \fref{fi:Post_pred_strain} we see that the inferred PDFs are able to reproduce the experimental strain values well, proving that inferred PDFs are indeed representative of the data. 

\begin{figure}[h]
  \centering
  \includegraphics[width=\columnwidth]{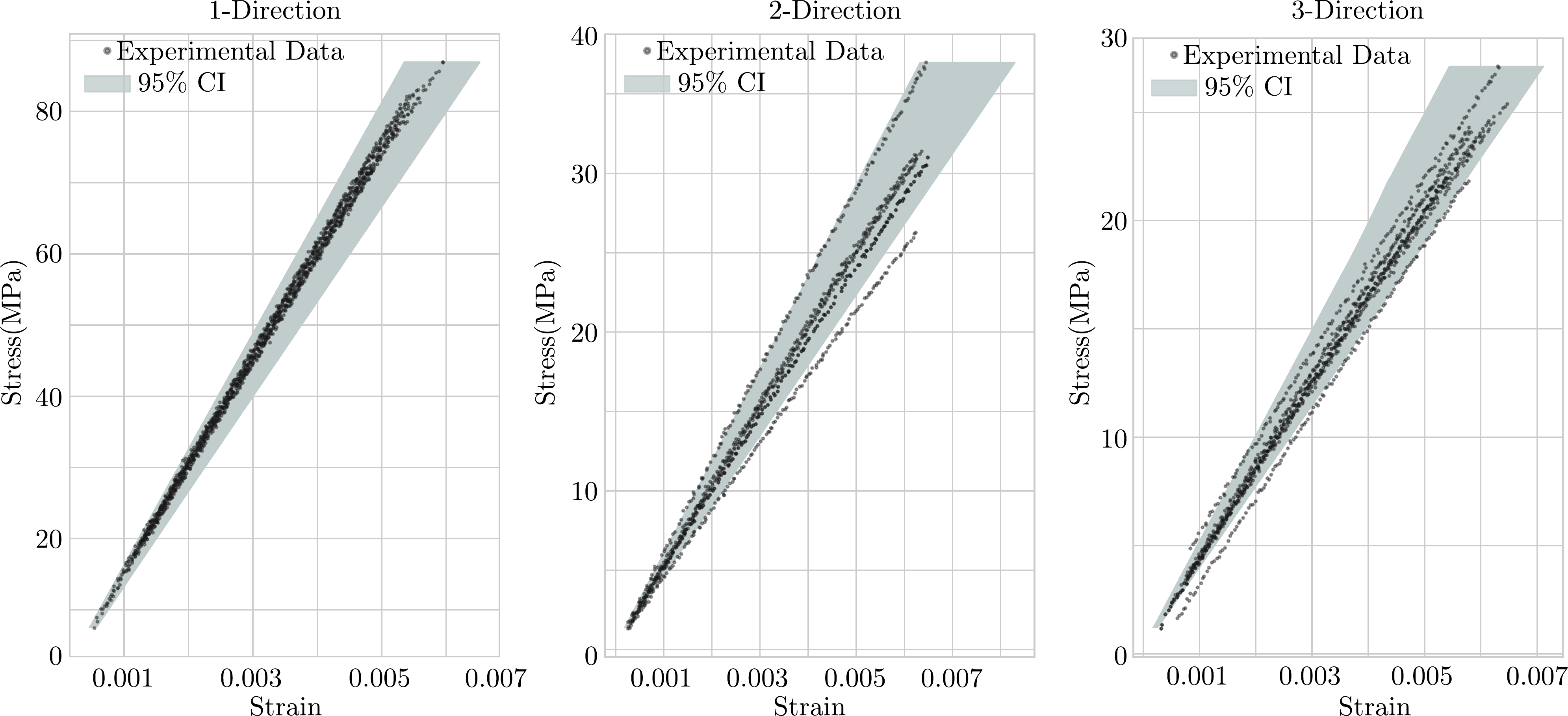} % scale=0.35
  \caption{Stress-strain experimental data vs model predictions.
  }
  \label{fi:HPD_PLOTS_STRAIN}
\end{figure}

However, it is more intuitive to analyze credible intervals predicted by the model and compare them to the stress strain data for the tensile tests in the three material directions. 
Credible intervals (CI) characterize the posterior of a random variable with point values.
Formally, a credible interval of size $(1-\alpha)$ (for the parameters $\Theta$ given the data $\textbf{X}$) in an interval $(a,b)$ is defined as:
\begin{equation}\label{credible interval}
  \int_a^b p(\Theta |\textbf{X}) d \Theta = 1 - \alpha. 
\end{equation}
In short, a credible interval $(a,b)$ contains $(1-\alpha)$\% of the posterior mass. 
However, credible intervals are not unique since there are infinitely many intervals $(a,b)$ that can contain $(1-\alpha)$\% of the posterior mass. 
Therefore, we use the highest posterior density interval (HDI). 
HDI is defined as the shortest interval that contains $(1-\alpha)$\%  of the posterior mass. 
We choose a value of $\alpha = 0.05$ in this work. 
\fref{fi:HPD_PLOTS_STRAIN} compares the model predicted strains with the 95\% HDI (denoted as 95\% CI) to the experimental data. 
From \fref{fi:HPD_PLOTS_STRAIN} we see that the reproduced strain data using the inferred properties are in extremely good agreement with the experimental strain data. 
Few data points lie outside the 95\% CI but there is excellent agreement overall. 
This supports the validity of the inferred PDFs. 
Further, we can make some interesting observations from \fref{fi:HPD_PLOTS_STRAIN}. 
We can note that the 95\% CI predicted by the Bayesian model overestimates the variation of experimental strain data in the 1-direction.
This is due to the fact that we used data from the 1-, 2-, and 3-directions simultaneously for the inference.
Therefore, the inferred properties show variation to match the strain data from all the three directions. 
It can be seen from \fref{fi:HPD_PLOTS_STRAIN} that the experimental variation of strain in the 2-direction and 3-direction are well predicted by the model. 
However, as a consequence, larger variation is predicted in the 1-direction as compared to the experimental data.

\begin{figure}[h]
  \centering
  \includegraphics[width = 0.9\linewidth]{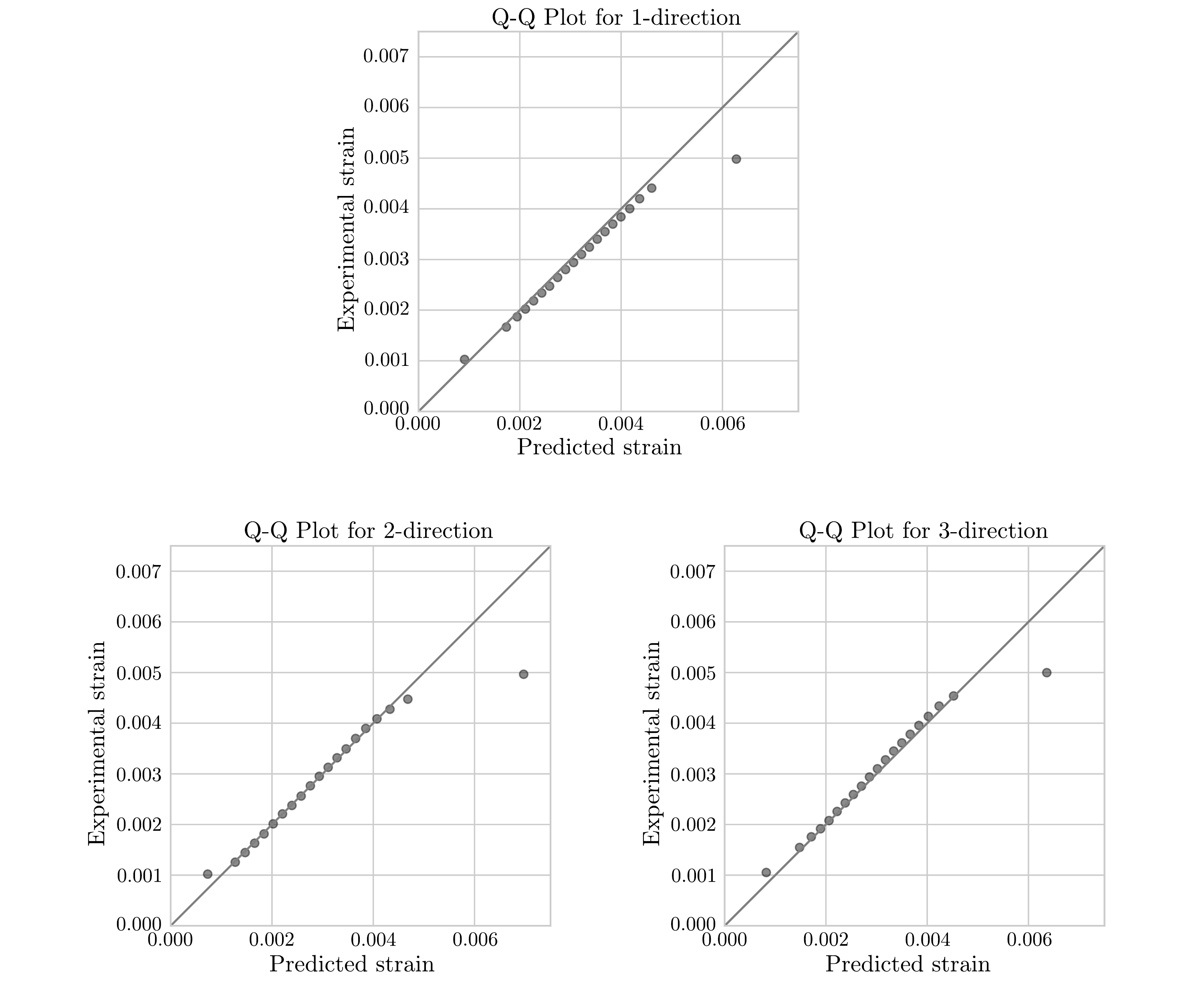} % scale=0.45
  \caption{Quantile-Quantile (Q-Q) plots }
  \label{fi:QQ_plot}
\end{figure}

Next, to compare the statistics of the experimental data to the statistics of the Bayesian predictions we analyze the quantile-quantile plot between samples from the predictive distribution and the experimental data. 
The distributions were split into 30 quantiles and the comparison is shown in \fref{fi:QQ_plot}. 
From \fref{fi:QQ_plot} the predictive quantiles closely match the experimental quantiles indicating the Bayesian predictions indeed represent the experiment data closely.

\begin{figure}[h!]
  \centering
  \includegraphics[width=0.85\linewidth]{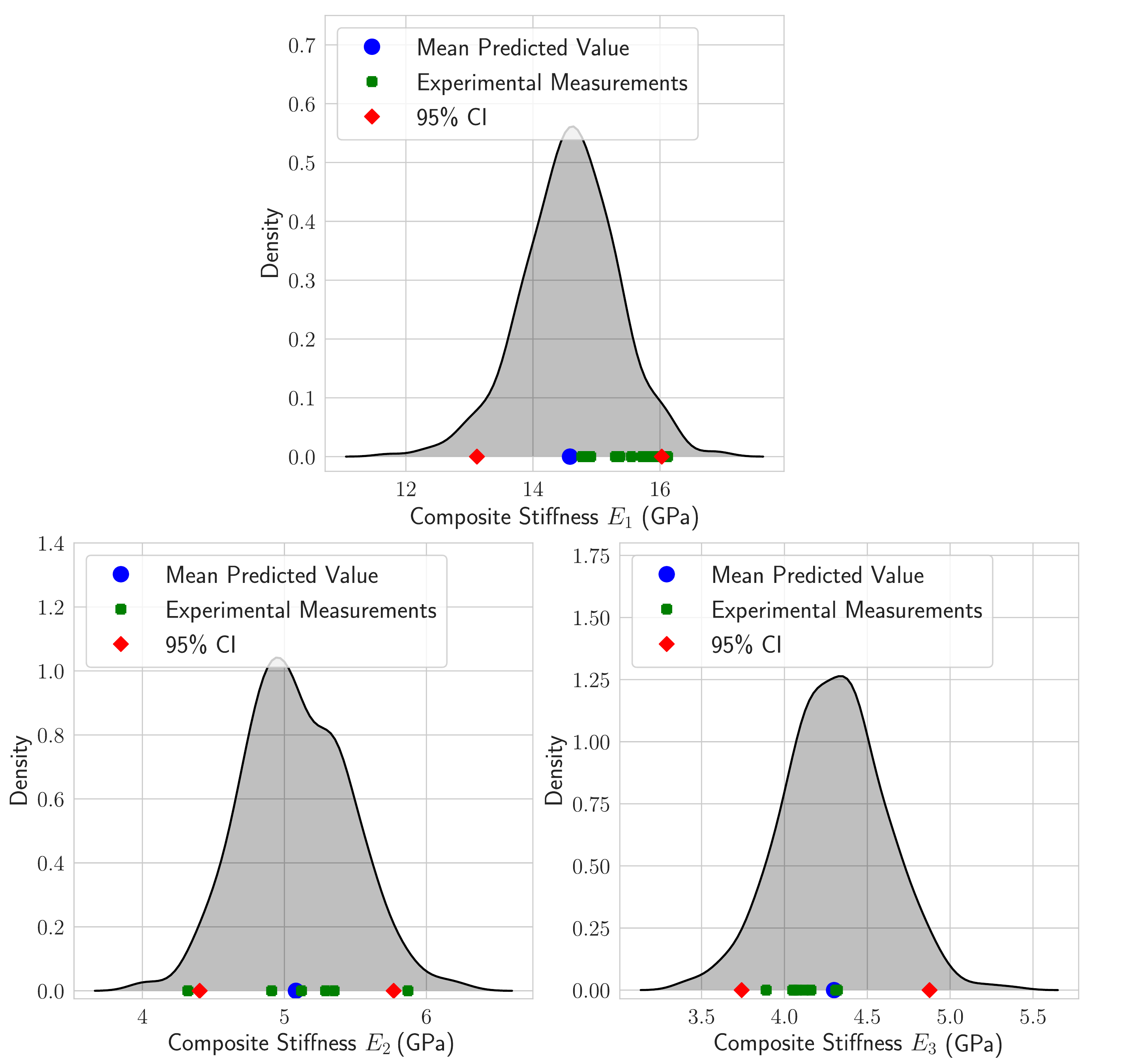} %scale = 0.4
  \caption{Predicted composite stiffness in the three directions}
  \label{fi:Stiffness_Credible_intervals}
\end{figure}

Next, we can generate a 95\% CI for the composite tensile stiffness values in the three directions using the $\GP$ surrogate. 
This is shown in \fref{fi:Stiffness_Credible_intervals}.
We also plot the different experimentally measured tensile stiffness values in \fref{fi:Stiffness_Credible_intervals}.
Note that we calculated the stiffness from the experimental stress strain data in accordance with  ASTM D3039 standard~\cite{standard2008standard}.
From \fref{fi:Stiffness_Credible_intervals}, we see that in the case of the 1-direction stiffness, most of the experimental measurements lie between the mean predicted value and the upper credible interval with just one measured stiffness slightly beyond the upper credible  interval.
In the case of the 2-direction stiffness, we see much wider spread in experimental measurements.
This variation is well captured in the predicted credible intervals as shown in \fref{fi:Stiffness_Credible_intervals}.
There is an instance where the experimentally measured stiffness is greater than the upper credible interval and lower than the lower credible interval for the 2-direction.
However, there is good agreement overall. 
In the case of the 3-direction stiffness predictions, we see that all the experimentally measured stiffness values are within the 95\% CI.

Next, we estimate the distribution of the nine independent constants of the composite elasticity tensor using the inferred properties. 
Specifically, we understand the effects of the uncertainty of our inferred parameters on the composite elasticity tensor components, or more commonly known as the uncertainty propagation (UP) problem.
To do so, we evaluate the micromechanics model discussed in \sref{se:micro_model} using the posterior samples of the matrix modulus and orientation values.
Note, we are able to estimate the distribution of the elasticity tensor components since the micromechanics model is not prohibitively expensive to evaluate. 
In the case that the micromechanics model is computationally expensive, approximating the prediction of each independent elasticity tensor component using a cheap-to-evaluate surrogate can overcome the computational bottleneck of the UP problem~\cite{tripathy2016gaussian}.

\begin{figure}[t!]
  \centering
  \includegraphics[width=\linewidth]{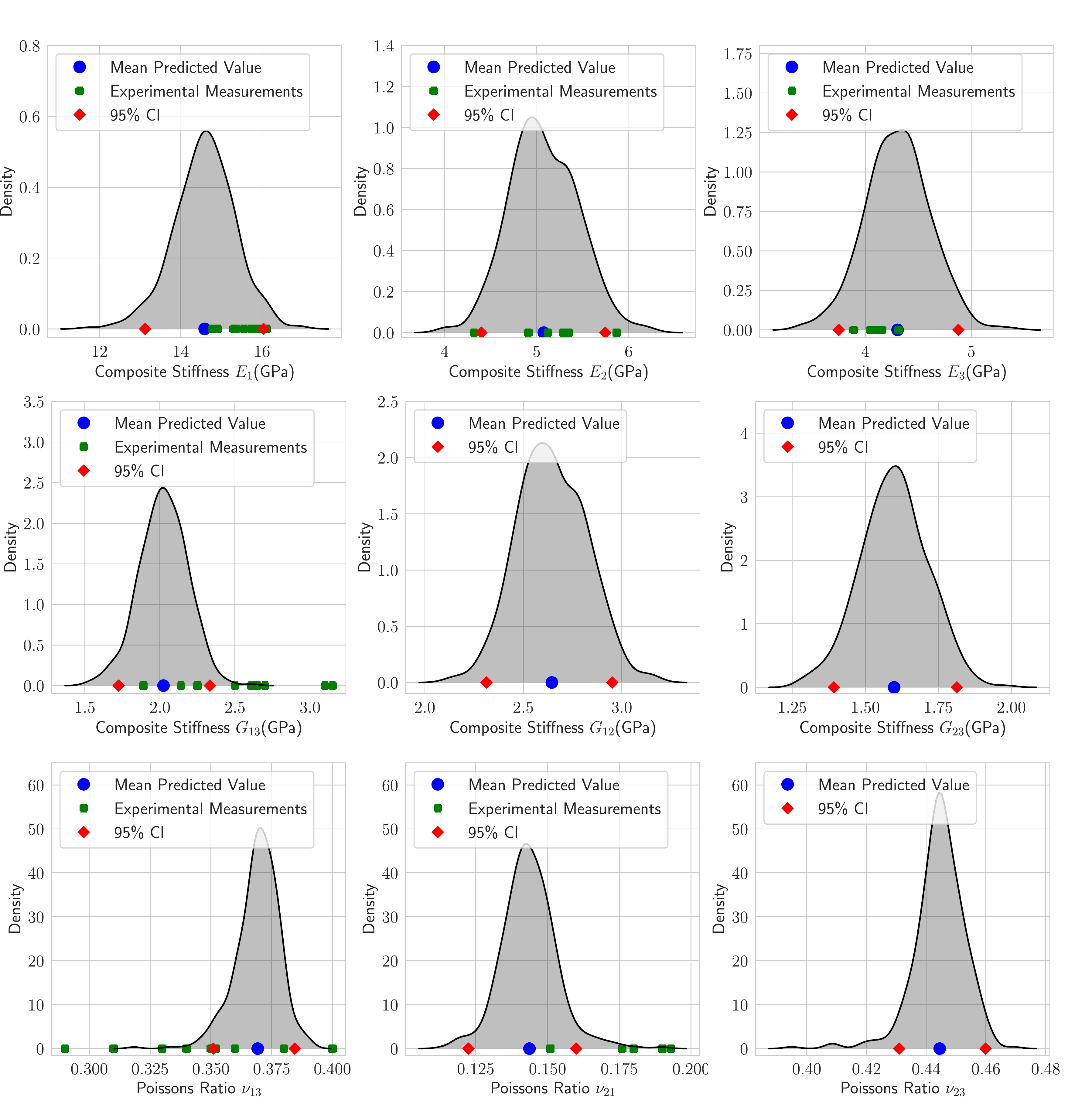} %scale = 0.4
  \caption{Composite elasticity tensor components with credible intervals}
  \label{fi:elastic_tensor_Credible_intervals}
\end{figure}

The distributions of the composite elasticity tensor predictions are shown in \fref{fi:elastic_tensor_Credible_intervals}.
The first row in \fref{fi:elastic_tensor_Credible_intervals} show the tensile stiffness prediction compared to the experimental measurements. 
We see that there is excellent agreement of the predicted values and the experimentally measured values, with most of the experimental values lying within the credible interval of the predictions.
This is expected since we use the data from these experiments for the inference.

The second row of \fref{fi:elastic_tensor_Credible_intervals} shows the distributions of the shear stiffness predictions of the composite. 
The shear modulus, $G_{13}$, was experimentally characterized according to the D5379 Standard Test Method for Shear Properties of Composite Materials by the V-Notched Beam Method \cite{astm20055379}. 
Comparing the $G_{13}$ predictions to the experimental measurements, we see that there is a bias in our predictions. 
This bias is evident by comparing the mean predicted values to the mean experimental values as shown in table  \ref{table:Bayesian_Composite_Predictions}.
From \fref{fi:elastic_tensor_Credible_intervals} we see that the some experimentally measured $G_{13}$, fall outside the predicted distribution.
While some data points lie within the predicted CI there are points that lie beyond CI.
This discrepancy could be attributed to the fact that the micromechanics model used does not account for the variation of orientation of the fibers within a printed bead. 
The effect of this variation could be pronounced under shear loading leading to different shear modulus predictions from the micromechanics model. 
Further, it should be noted that the fiber properties were taken~\cite{king1992micromechanics} which might not represent the actual fiber used in this SFRP system. 

Next, the third row of \fref{fi:elastic_tensor_Credible_intervals} shows the distributions of the Poisson ratio predictions of the composite. 
Comparing the predicted distribution of $\nu_{13}$ to the experimentally measured values, we see that the experimental measurements of $\nu_{13}$ show a wider spread than the predictions. 
Experimental noise could be a factor contributing to this wide spread.
However, there is minimal bias in the predictions compared to the experimental measurements of $\nu_{13}$. 
This can be seen by comparing the mean predicted values to the mean experimental values of $\nu_{13}$ as shown in table \ref{table:Bayesian_Composite_Predictions}. Finally, by comparing the predicted distribution of $\nu_{21}$ to the experimental measurements we observe a bias in the predictions. Most of the experimental measurements lie outside of the predicted CI. 
The discrepancy in the $\nu_{21}$ predictions could be aggravated by  the fact that signal-to-noise ratio is low when measuring small strains via the DIC.
To clarify, the strain in the 1-direction (printing direction) is low due to the preferential alignment of fibers in this direction during the printing process.
This could lead to erroneous experimental measurements and thereby augmenting the prediction error. 

Our results indicate that we can populate the elasticity tensor of the composite, with the uncertainties quantified, with as few as three tensile tests. 
These uncertainties can be propagated through manufacturing simulation codes to obtain statistics of a quantity of interest, e.g., mean and variance of deformation predictions, thereby enabling a step towards a probabilistic predictive framework for composites manufacturing digital twins.

\begin{table}[h!]
\caption{Mean composite property predictions vs mean experimental measurements} % title of Table
\centering % used for centering table
\begin{tabular}{| c || c | c |} % centered columns (3 columns)
\hline\hline %inserts double horizontal lines
Property &  Mean Predictions & Mean Experimental Measurements\\ [0.7ex] % inserts table
%heading
\hline % inserts single horizontal line 
$E_1 (GPa)$ & 14.65 & 15.42\\ % inserting body of the table
$E_2 (GPa)$ & 5.02 & 5.14\\
$E_3 (GPa)$ & 4.33 & 4.12\\
$G_{12} (GPa)$ & 2.61 & Not Measured\\
$G_{13} (GPa)$ & 2.04 & 2.55\\ 
$G_{23} (GPa)$ & 1.59 &  Not Measured\\
$\nu_{21} $ & 0.14 & 0.18\\
$\nu_{13} $ & 0.37 & 0.35\\
$\nu_{23} $ & 0.45 & Not Measured\\
\hline %inserts single line
\end{tabular}
\label{table:Bayesian_Composite_Predictions} % is used to refer this table in the text
\end{table}

\section{Conclusions}
\label{se:conc}

% In summary, this work presented a Bayesian framework to infer the orientation of the fibers and the modified polymer properties in a SFRP composite. This study was motivated by the need to reduce extensive experimental efforts to measure the fiber orientation and modified polymer properties.  In particular, the EDAM process was used to study a 20 percent by weight CF-PEI short-fiber material system. The material was printed on the CAMRI system from which the orientation tensor and the modified polymer properties were inferred. The epistemic and aleatory uncertainties were also quantified in the Bayesian inference framework. Statistical tests showed that the inferred parameter distribution were able to recreate the observed experimental closely.

The methods discussed in this work enable the user to infer composite properties in discontinuous fiber composites with as few as three tensile tests. 
The hierarchical Bayesian approach quantifies the uncertainties associated with the material properties by making use of the stress-strain data as opposed to the average stiffness of the composite. 
The inferred parameter distribution are the distributions across the different specimens tested rather than the distributions within a specimen. 
This is a result of the fact the strain was averaged over the region of interest (shown in \fref{fi:DIC}) while measuring strain using DIC. 
Since the focus of this work is to populate the elasticity tensor of a short fiber composite, we have not examined various sizes of regions of interest for DIC strain measurements.
However, it can affect the inferred property distributions and should be a topic of further investigation. 
Further, the predicted strain distribution using the inferred properties overestimated the strain variation in the 1-direction. 
This overestimation was a consequence of using strain data in the three directions, i.e. the inferred parameter distributions were such that it reflected the wide variation of strain the 2- and 3-directions. 
As a result, the predictions showed more strain variation in the 1-direction compared to the experimental data. 
Although, we use the two step homogenization process, our development is such that it can be extended to more complex homogenization methods, e.g. finite element based homogenization techniques. This flexibility is enabled by replacing the micromechanics model using an inexpensive Gaussian process ($\GP$) surrogate. 
We develop the GP surrogate such that a different micromechanics models can be used to predict the composite stiffness in each direction. 
Since the micromechanics model used in this work was not prohibitively expensive, we could create sufficient data to safely ignore the uncertainties associated with the surrogate predictions (predictive variance was negligible). 
When data is not abundant, we cannot ignore such uncertainties.

Nonetheless, the potential of the presented framework is in quantifying uncertainties associated in the inference process and predicting the uncertainties associated in the elasticity tensor of the composite. 
These calibrated uncertainties can be propagated through simulation codes to obtain statistics of a quantity of interest, (e.g., mean and variance of deformation predictions) and thereby is of immense importance in enabling a probabilistic framework for composites manufacturing digital twins. 
In light of the current limitation of the method, we propose the following directions for future work. 
We propose to use an active learning approach to develop micromechanics surrogates, since creating a large dataset will be computationally prohibitive when using expensive mircomechanics models. 
The resulting uncertainties associated with the surrogate predictions due to the limited dataset will be incorporated in the inference of the matrix modulus and the fiber orientation. 
Further, the predicted elasticity tensor uncertainties should be propagated through composite manufacturing simulation codes to predict deformations of the resulting part. This should be compared to experiments to understand the effect of the composite elasticity tensor uncertainties.

\section*{CRediT author contribution statement}
\textbf{Akshay J. Thomas}: Conceptualization, Methodology, Software, Investigation, Formal analysis, Writing - original draft. \textbf{Eduardo Barocio}: Conceptualization, Investigation, Writing - review and editing. \textbf{Ilias Bilionis}: Methodology, Investigation, Formal Analysis, Writing - review and editing. \textbf{R. Byron Pipes}:  Investigation, Funding acquisition, Supervision, Resources, Writing - review and editing.

\section*{Declaration of competing interest}
The authors declare no competing financial or personal interests that could have appeared to influence this work.

\section*{Acknowledgements}
This research was sponsored by the U.S. Department of Energy, Institute for Advanced Composites Manufacturing Innovation (IACMI), under contract DOE DE-EE0006926, IACMI PA16-0349-3.12-01.

% \section*{Acknowledgements}
% Thank your project partners or people who helped contribute to the work.
\fontsize{10}{10}\selectfont
%\clearpage 
\bibliographystyle{model1-num-names}
\bibliography{references.bib}

\end{document}